 \documentclass[final,3p,times,twocolumn]{elsarticle}
\usepackage{amssymb}

\journal{Astroparticle Physics}

\begin{document}

\begin{frontmatter}


 \ead{maria.giller@kfd2.phys.uni.lodz.pl}

\title{An analytical approach to the multiply scattered light in the optical images of the extensive air showers of ultra-high energies}


\author{Maria Giller and Andrzej \'Smia\l{}kowski}

\address{The University of Lodz, Department of High Energy Astrophysics, Pomorska 149/153, 90-236, Lodz, Poland}

\begin{abstract}
One of the methods for studying the highest energy cosmic rays is to measure the fluorescence
light emitted by the extensive air showers induced by them. 
To reconstruct a shower cascade curve from measurements of the number of photons arriving from
the subsequent shower track elements it is necessary to take into account the multiple scatterings 
that photons  undergo on their way from the shower to the detector. 
In contrast to the earlier Monte-Carlo work, we present here an analytical method to  
treat the Rayleigh and Mie scatterings in the atmosphere. 
The method consists in considering separately the consecutive 'generations' of the scattered light. 
Starting with a point light source in a uniform medium, we then examine a source in a real atmosphere and finally - a moving source (shower) in it.
We calculate the angular distributions of the scattered light superimposed on the not scattered light registered from a shower at a given time. The analytical solutions (although approximate) show how the exact numerical results should be parametrised what we do for the first two generations (the contribution of the higher ones being small). Not allowing for the considered effect may lead to an overestimation of shower primary energy by $\sim 15\%$ and to an underestimation of the primary particle mass.

\end{abstract}

\end{frontmatter}
\section{Introduction}

One of the methods for studying extensive air showers of high energies ($\ge 10^{17}$ eV) is to register their images in the optical (mainly  fluorescence) light. This can be done by observing showers from the side in order to avoid the more intense Cherenkov light emitted rougly in the shower direction. The observations are made by a number of optical telescopes, each containing a large mirror and a camera with a matrix of photomultipliers (PMTs) placed at the focus of the optical system (HiRes \cite{hires}, The Pierre Auger Observatory \cite{pao}, The Telescope Array \cite{tarr}), so that photons arriving from a given direction on the sky are focused on a particular PMT (pixel). Photon arrival time can also be measured  if the time structure of  the PMT signals is recorded.
 A cosmic ray induced shower produces at a given time a light spot on the camera which moves across it  as the shower develops in the atmosphere  so that  succeeding PMTs are being hit. \\
It would have been ideal if the light producing a shower  image had contained the fluorescence photons only. This is because there  exists experimental evidence that the number of  fluorescence photons induced by a charged electron in the atmosphere  is proportional to the energy lost by it for ionisation \cite{dedx}. As practically all primary particle energy is eventually used for ionisation, this energy can be determined by measuring the fluorescence light emitted along the shower track in the atmosphere.\\
However, there are several problems in deriving the flux of the fluorescence light emitted by a shower from that  arriving at the detector.
Firstly, the arriving light contains not only the fluorescence  but also  Cherenkov photons. If the viewing angle (the angle between the line of sight and the shower direction) is large (say, $> 30^\circ$) then it is mainly the  Cherenkov  light \emph{scattered}  in the atmosphere region just passed  by  the shower  and observed by the detector (this light, before being scattered, travels roughly along the directions of the shower particles). Typically its fraction at the detector is about $15\%$ of the fluorescence flux. For smaller viewing angles it is the Cherenkov light \emph{produced} at the observed part of the shower that may dominate even the fluorescence signal. The contribution of the Cherenkov light, which has to be subtracted from the total signal, has been extensively studied \cite{espec,simil,nerl,latch}.\\
The subject of this paper is another phenomenon, affecting shower images, most commonly called the multiple scattering (MS) of light. Photons produced at the observed shower element, whatever their origin (fluorescence or Cherenkov),  may undergo scattering in the atmosphere on their way from the shower to the detector, causing an attenuation of the light flux arriving at the detector and a smearing of the image. This scattering may take place on the air molecules (Rayleigh scattering) or on larger transparent particles, aerosols (Mie scattering). Most of the scattered photons change their directions, so that they no longer arrive at the pixel registering the not scattered (direct) light. Moreover, they arrive later having longer path lengths to pass. On the other hand, photons  emitted by the shower at earlier times and scattered somewhere, may fall in the field of view of the pixels  just registering the direct photons emitted at a later time. The net effect is that  the scattered light forms its own instantaneous image superimposed on  that in the direct light.\\
  Our aim is to calculate the shower images in the multiply scattered light, so that this effect could be allowed for (subtracted) when determining the shower primary energy from the PMT signals. This problem was already studied by Roberts \cite{roberts}, by us in several, short conference contributions \cite{giller} and more recently by P\c ekala et al \cite{pekala}. The approach of the other authors was based on Monte Carlo simulations  of    
 photons emitted by a shower. Photons were followed up to 5-6 scatterings and their arrival directions and time were registered by the detector. Many shower simulations were needed to obtain the MS images for various distances, heights, viewing angles of the observed shower parts. Finally, a phenomenological parametrisation of the number of MS photons was made as a function of the parameters found as relevant.\\
In contrast to Roberts and P\c ekala et al  this approach is based on an analytical treatment. The main idea is to consider the arriving MS light as a sum of the photons scattered only once (the first generation), of  those scattered two times (the second generation) and so on and calculate separately the angular and temporal distributions for each generation.\\
We  start (Section 2) with a consideration of the simplest situation when a point source of isotropic light flashes for a very short time in an uniform medium. We derive analytical expressions for the angular and temporal distributions of the first and next generations of light arriving at a particular distance from the source.  \\
As our aim is to apply our results to  cosmic ray showers we need to consider a non-uniform medium like the atmosphere. Assuming an exponential distribution of the gas density and similarly for aerosols we show that an effective scattering length between any two points in the atmosphere can be easily calculated analytically. Signals of the first two generations arriving at a particular detector within a given angle $\zeta$ to the direction to the source are found as a function of time (Section 3).\\
Using these it is straightforward to derive the corresponding distributions if the source moves across the atmosphere, integrating the point source distributions over changing distance and time of light  emission.
In Section 4 we consider a moving light source, modelling a distant shower. We calculate angular distributions of MS light arriving at a detector at the same time as the direct (not scattered) photons emitted by the shower. This particular approach is quite natural because the data from optical detectors  consist of the recorded signals by the camera PMTs within short time intervals $\Delta t$  so that one needs to know how much of the MS light has to be subtracted from the main, direct  signal. The method for calculating images in the MS light \emph{simultaneous} with images in the direct light is relatively simple (for the first and the second generations) as it is based on the geometry  of the scattered photons in the particular generation. It does not require time-consuming Monte-Carlo simulations that were done for various shower-observer geometries.
Making some approximations we derived analytical formula for a shower image produced by the first generation (Section 4.2). Our analytical derivations allowed us to choose easily the variables on which and how to parametrise the MS signals. They also made us to realise that the dependence of it on the viewing angle was different for Rayleigh  and Mie scatterings. Thus, we introduced a new, simple parametrisation of the fraction of the scattered photons arriving at the telescope within a given viewing cone, depending on the viewing angle of the shower (what has not been done before), separately for the two scatterings. For the same reason we also parametrised separately the second generation (Section 4.3). \\
A discussion of the results and the implication of the MS effect on the derivation of shower parameters is given in Section 5. The last Section (6) contains a summary and conclusions.
\section{Point source flashing isotropically in uniform medium}
At any fixed time a distant shower can be regarded as a point source emitting isotropically fluorescence light (about Cherenkov light see later). As explained above, we treat the light scattered in the medium as a sum of consecutive generations consisting of photons scattered only once, twice, and so on, on their way from the source to the observation point.
\begin{figure}[th]
  \centering
  \includegraphics[width=3in]{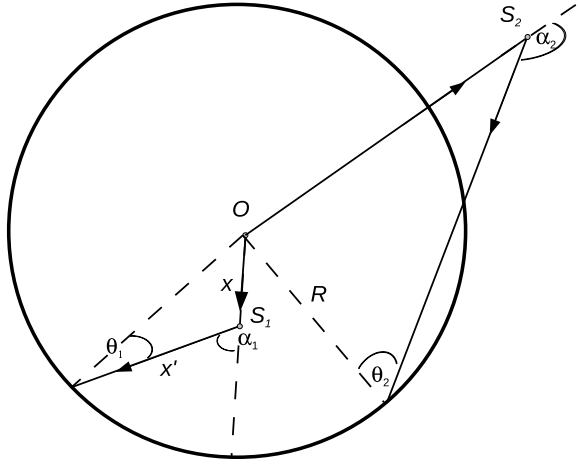}
  \caption{ Geometry of the first scattering in a uniform medium. The light source is at the centre $O$ of a sphere with radius $R$. Two rays shown are scattered at points 
$S_1$ and $S_2$, correspondingly, arriving at the surface of the sphere with radius $R$ at angles $\theta_1$ and 
$\theta_2$.}
  \label{fig1}
\end{figure}
\subsection{First generation}
Let us consider the first generation, consisting of photons scattered once only. We shall calculate the flux of these photons, $j_1(\theta, t; R)$, at a distance $R$ from the source, such that $j_1(\theta, t; R)d\Omega dt dS_\perp$ is the number of photons scattered only once, arriving at time $(t, t+dt)$ after the flash, within a solid angle $d\Omega (\theta)$ at the surface $dS_\perp$ (perpendicular to the arrival direction) located at a distance $R$.
To do this we shall calculate first the number of photons crossing the sphere of radius
$R$ (from inside) at an angle ($\theta, \theta+d\theta$) with respect to the normal, at time ($t, t+dt$); see Fig. 1.
 The average number of photons, per one photon emitted, interacting at a distance ($x, x+dx$) from the source and scattered at an angle $\alpha$ within 
$d\Omega(\alpha)$, equals
\begin{equation}
d n_1(x,\alpha)=e^{-\frac{x}{\lambda}}\frac{dx}{\lambda}f(\alpha)d\Omega(\alpha)\cdot e^{-\frac{x'}{\lambda}}
\end{equation}
where $\lambda$ is the mean scattering path length, $f(\alpha)d\Omega$
is the probability that, once the scattering has occured, the scattering angle is $\alpha$ within 
\mbox{$d\Omega(\alpha)=2\pi \, sin\alpha\, d\alpha$}. 
To each pair of variables $(x, \alpha)$ there corresponds another pair $(\theta, t)$ related to the former by
\begin{equation}
tg \frac{\alpha}{2}=\frac{\tau-cos\theta}{sin\theta}
\end{equation}
and
\begin{equation}
x=\frac{R}{2}\,\frac{\tau^2-2\tau\, cos\theta+1}{\tau-cos\theta}
\end{equation}
where $\tau=ct/R$ and $c$ is the speed of light.\\
The Jacobian of the transformation gives
\begin{equation}
sin\alpha\,d\alpha\, dx=\frac{2Rcos\theta}{\tau^2-2\tau \,cos\theta+1}d\tau|d\, cos\theta|
\end{equation}
Thus, we obtain
\begin{equation}
d n_1(\theta, t)=4\pi\frac{c}{\lambda}e^{-\frac{ct}{\lambda}}\frac{f(\alpha)sin\theta\,cos\theta}{\tau^2-2\tau \,cos\theta+1}d\theta\,dt
\end{equation}
Finally, the number of photons arriving at a unit surface at an angle $(\theta, \theta+d\theta)$
(all azimuths) at time $(t, t+dt)$ equals
\begin{equation}
d n_1(\theta, t)=\frac{c}{\lambda\,R^2}\frac{e^{-\frac{ct}{\lambda}}\,f(\alpha)sin\theta\,|cos\theta|}{\tau^2-2\tau \,cos\theta+1}d\theta\,dt
\end{equation}
and
\begin{equation}
j_1=\frac{1}{2\pi sin\theta\,|cos\theta|}\frac{d^2n_1}{d\theta dt}
\end{equation}

\subsubsection{Rayleigh scattering}
For the Rayleigh scattering we have
\begin{equation}
f(\alpha)=f^R(\alpha)=\frac{3}{16\pi}(1+cos^2\alpha)
\end{equation}
Expressing $\alpha$ as a function of $\theta$ and $\tau$ (Eq. 2) we obtain
\begin{equation}
f^R=\frac{3}{8\pi}\Big[1-\frac{2sin^2\theta}{y}+
\frac{2sin^4\theta}{y^2}\Big]
\end{equation}
where $y=\tau^2-2\tau\, cos\theta+1$.
Thus, the flux $j_1^R(\theta, t; R)$ of the first generation, defined above, equals
\begin{eqnarray}
j_1^R(\theta, t; R)= \frac{3c\,e^{-\frac{ct}{\lambda}}}{16\pi^2\lambda_RR^2\,y}\cdot
\Bigg(1-\frac{2sin^2\theta}{y}+\frac{2sin^4\theta}{y^2} \Bigg) \nonumber \\
\,
\end{eqnarray}
where $\lambda_R$ is the mean free path length for the Rayleigh scattering. However,
$\lambda$ in the exponent  depends on all the scattering processes active. In general it is determined by 
\begin{equation}
\frac{1}{\lambda}=\sum_i^n\frac{1}{\lambda_i}
\end{equation}
for $n$ processes. Thus, if both molecular and aerosol scatterings are active but one wants to calculate the flux of photons scattered by the  Rayleigh process only, 
$\lambda$ in the exponent equals $\lambda=(\frac{1}{\lambda_R}+\frac{1}{\lambda_M})^{-1}$
but in the denominator one has $\lambda_R$.\\
From Eq. 10 one can find the number of photons $\frac{d N_1^R(t;\zeta,R)}{dt}$ arriving per unit time at a unit surface within a given angle $\zeta$, as a function of time.\\
We have
 \begin{eqnarray}
\frac{d N_1^R(t;\zeta,R)}{dt}=\int_0^{\zeta}j_1^R(\theta, t; R)\cdot 2\pi \, sin\theta|cos\theta|d\theta \nonumber\\
\,
\end{eqnarray}
The integral can be found analytically, giving the result
\begin{eqnarray}
\lefteqn{\frac{d N_1^R(t;\zeta,R)}{dt}=}\nonumber \\
&&=\frac{3ce^{-\frac{ct}{\lambda}}}{8\pi \lambda_R R^2}\cdot \frac{1}{32\tau^6}
\Bigg(\big(\tau^2+1\big)\frac{a_{-2}}{2} \big(\frac{1}{y_2^2}-\frac{1}{y_1^2}\big)+ \nonumber \\
&&+ \sum_{i=-2,0,1} \Big[(\tau^2+1)a_{i+1}-a_i\Big]\cdot\frac{y_2^{i+1}-y_1^{i+1}}{i+1}+ \nonumber \\
&&+ \Big[ (\tau^2+1)a_0-a_{-1} \Big] ln \frac{y_2}{y_1}-a_2 \frac{y_2^3-y_1^3}{3} \Bigg)
\end{eqnarray}

where $y_1=(\tau-1)^2$, $y_2=\tau^2-2\tau\, cos\zeta+1$, $a_2=1$, $a_1=-4$, $a_0=6(\tau^4+1)-4\tau^2$, $a_{-1}=-4(\tau^2-1)^2$,
$a_{-2}=(\tau^2-1)^4$.\\
We have also calculated analytically a similar distribution $dN_1^{is}/dt$ if the scattering was isotropic. i.e. if $f(\alpha)=\frac{1}{4\pi} $ (Appendix A).\\
One can also find analytically the angular distribution $\frac{d M_1^R}{d\theta}$ of the arriving light (integrated over time), but for small angles only (Appendix B). The result is
 \begin{eqnarray}
\lefteqn{\frac{d M_1^R(\theta;R)}{d\theta}=2\pi sin\theta\, cos\theta \int_{R/c}^{\infty} j_1^R(\theta,t;R)dt=} \nonumber\\
&=&\frac{9k_Re^{-k_R}}{64R^2}\Bigg( 1-\frac{4\theta}{3\pi}+
\frac{8k_R\theta }{3\pi}\Big[ ln(k_R\theta)+C_{Eu}-\frac{1}{2}\Big] \Bigg) \nonumber\\
\,
\end{eqnarray}
where $\theta\ll 1$, $k_R=\frac{R}{\lambda_R}$, if there is no Mie scattering and $C_{Eu}\simeq0.577$ is the Euler constant.\\
The ratio of all photons arriving within a small angle $\theta$ to those not scattered $N_0$, equals
 \begin{eqnarray}
\lefteqn{\frac{1}{N_0}\int_0^{\theta}\frac{d M_1^R}{d\theta'}d\theta'\simeq
\frac{4\pi R^2}{e^{-k_R}}\frac{9}{64}\frac{k_Re^{-k_R}}{R^2}\theta=}\\
&=&\frac{9\pi}{16}k_R\, \theta(rad)\simeq 3.1\cdot10^{-2}k_R\,\theta (deg)\nonumber
\end{eqnarray}
where terms $\sim\theta^2$ have been neglected.
\subsubsection{Mie scattering}
In the next paragraph we shall consider the scattering of light emitted by showers developing in the real atmosphere, i.e. with the density depending on height. In addition to the Rayleigh process one has to take into account the Mie scattering occurring on particles (aerosols) larger that the light wavelength.
The Mie angular distribution is concentrated at rather small angles, in contrast to the Rayleigh case. Moreover, 
in the deeper parts  of the atmosphere the mean free path length for the Mie scattering may be comparable to that for Rayleigh, so that it is necessary to calculate the distribution of the light scattered by the Mie process only.\\
As before, we start with a simpler case - a uniform medium. The angular distribution of light scattered on particles with sizes larger
than the light wavelength depends on the distribution of the sizes and is not a well known function. Roberts \cite{roberts} adopts a function of the form 
\begin{equation}
f(\alpha)\sim e^{-B\alpha}+Ce^{D\alpha}
\end{equation}
Here, however, we prefer an expression allowing us to perform some integrations analytically. Most 
crucial is to have the number of numerical integrations for the second generation as few as possible.
We shall see that to find $j_2^R(\theta, t; R)$ for the Rayleigh scattering (Section 2.3) there is only one integration (over $x'$) to be done numerically since the form of $f^R(\alpha')$ enables one to integrate analytically over $\phi'$ and $\theta'$ (Eq. 25 and 26). Thus, we adopt the following form for the Mie angular
distribution:
\begin{eqnarray}
f_1^M(\alpha)&=&a_1 cos^8\alpha+b \,\quad \emph{for} \,\quad 0\le\alpha\le\frac{\pi}{2} \nonumber \\
f_2^M(\alpha)&=&a_2 cos^8\alpha+b \,\quad \emph{for}\,\quad \frac{\pi}{2}\le\alpha\le\pi
\end{eqnarray}
where $a_1=0.857$, $a_2=0.125$, $b=0.025$.\\
This function is normalised as follows
\begin{equation}
\int_0^{\pi} f(\alpha)\cdot2\pi\, sin\alpha\,d\alpha=1
\end{equation}
It describes quite  reasonably the distribution used by Roberts.\\
From Eq. 6 and 7 we have
\begin{equation}
j_1^M(\theta,t; R)=\frac{c e^{-k\tau}\cdot f_i^M\big[\alpha(\theta, t)\big]}{2\pi \lambda_M R^2(\tau^2-2\tau\,cos\theta+1)}
\end{equation}
where $i=1$ if $tg\frac{\alpha}{2}=\frac{\tau-cos\theta}{sin\theta}<1$ \\
and $i=2$ if $tg\frac{\alpha}{2}>1$\\
Since 
\begin{equation}
cos\alpha=\frac{1-tg^2\frac{\alpha}{2}}{1+tg^2\frac{\alpha}{2}}=\frac{2\,sin^2\theta}{\tau^2-2\tau\,cos\theta+1}-1
\end{equation}
we obtain
\begin{equation}
j_1^M(\theta, t; R)=\frac{c\,e^{-k\tau}}{2\pi\lambda_MR^2\,y}\Bigg[a_i\Big( \frac{2\,sin^2\theta}{y}-1\Big)^8+b\Bigg]
\end{equation}
where $y=\tau^2-2\tau\,cos\theta+1$, and $i$ is determined as before. In principle, it is possible to find analytically
the number of photons $\frac{d N_1^M(t;\zeta,R)}{dt}$ arriving within an angle $\zeta$ after time $t$ per unit time. However, each of the nine integrals 
\begin{equation}
I_n=\int_0^{\zeta}\frac{1}{y}\Bigg(\frac{sin^2\theta}{y}\Bigg)^n sin\theta\, cos\theta\,d\theta
\end{equation}
contains many terms itself, so that an analytical dependence on $\zeta$ and/or $\tau$ would be practically lost. 
Thus, we have found $\frac{d N_1^M}{dt}$ by integrating the flux (Eq. 21) numerically (similarly to Eq. 12).\\
Finally, the total flux of the first generation is the sum of the two fluxes arising from the two active mechanisms
of the scattering
\begin{equation}
j_1(\theta, t; R)=j_1^R+j_1^M
\end{equation}

\subsection{The second generation}
 \begin{figure}[th]
  \centering
  \includegraphics[width=3in]{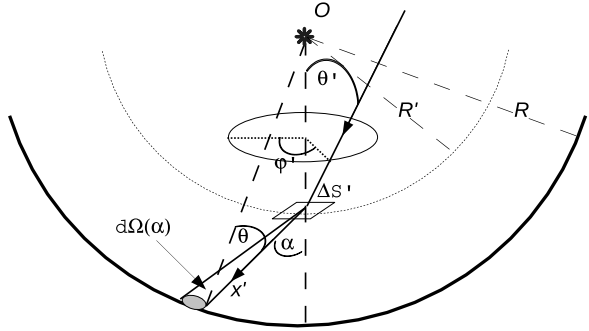}
  \caption{ Geometry of the second and higher scatterings in a uniform medium. The light source is at $O$ and the detector is the surface of the sphere with radius $R$. The picture shows the last scattering.}
  \label{fig2}
 \end{figure}

 \begin{figure*}[th]
  \centering
  \includegraphics[width=4in, angle=270 ]{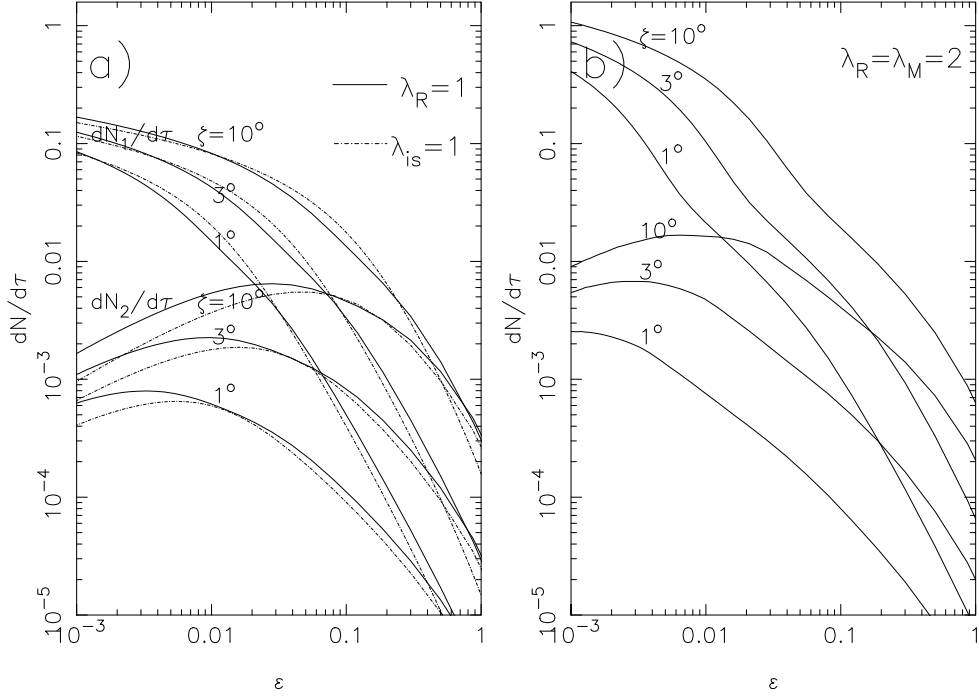}
  \caption{ Comparison of the first ($d N_1/d\tau$) (three upper curves) and the second ($d N_2/d\tau$) (three lower curves) generations as functions of time ($\epsilon=\tau-1=ct/R-1$). Number of photons are within angle $\zeta$. Uniform medium, $R=1$. \textbf{a).} Rayleigh (solid lines) and isotropic (dashed lines) scattering. 
\textbf{b).} Two scattering processes at work: Rayleigh and Mie, each with $\lambda=2R$.}
  \label{fig3}
 \end{figure*}

These are the photons scattered exactly two times. We shall consider first a general case when there are more than one scattering processes (as Rayleigh and Mie). Let us call the process of the first scattering as $A$ and this of the second one as $B$. Both $A$ and $B$ can be either Rayleigh or Mie. We denote their mean scattering path lengths by $\lambda_A$ and $\lambda_B$, and the angular distribution functions of the scattering by $f^A(\alpha)$ and $f^B\alpha)$, correspondingly. The light source flashes isotropically at the centre of a sphere with radius $R$ at time $t=0$ (Fig. 2). As before we want to calculate the number of photons crossing the surface of the sphere from inside at a given angle $\theta$, at time $t$, per unit time.\\
Let us consider the photons scattered for the second time at a distance $R'$ from the source. The number of photons incident on a small surface $\Delta S'$ at an angle $\theta'$ (within $d\Omega'$) at time $(t', t'+dt')$ and scattered within a spherical shell of thickness $dR'$ by an angle
 $\alpha'$ (within $d\Omega(\alpha')$) equals
\begin{equation}
j^A_1(\theta', t'; R')d\Omega'dt'\Delta S'|cos\theta'|\frac{dx}{\lambda_B}f^B(\alpha')d\Omega
\end{equation}
where $dx=dR'/|cos\theta'|$. As now both processes are active the meaning of $\lambda$ in the factor $e^{-\frac{ct'}{\lambda}}$ in the expression for $j_1^A$ is the effective mean path length for both processes.\\
The direction of the scattered photons is at an angle $\alpha$ to the radius of the sphere and $d\Omega=sin\alpha\,d\alpha\, d\phi$, where $\phi$ is the azimuth of the photons scattered for the second time. For any given direction $(\theta', \phi')$ before and $(\alpha,0)$ after the second scattering, the scattering angle $\alpha'$ fulfils the relation
$cos\alpha'=cos\alpha\, cos\theta'-sin\alpha\, sin\theta'\,cos\phi'$.
The only function depending on the azimuth angle $\phi'$ of the incident photons is $f^B(\alpha')$. Denoting
\begin{equation}
F^B(\theta', \alpha)=\int_0^{2\pi}f^B(\alpha')d\phi'
\end{equation}
and integrating (25) over $\theta'$ we obtain for the number of photons incident on $\Delta S'$ and scattered within $dR'$ into the solid angle $d\Omega(\alpha)$ the following expression
\begin{eqnarray}
\lefteqn{dt'\Delta S'dR'd\Omega\int_0^{\pi}j_1^A(\theta',t'; R') F^B(\theta',\alpha)sin\theta'd\theta' =} \nonumber \\
&&= G^{AB}(R',\alpha)\cdot dt'\Delta S'dR'd\Omega
\end{eqnarray}
where the function $G^{AB}(R',\alpha)$ is defined by the integral in the l.h.s. of (26).\\
The pair of fixed variables $R'$ and $\alpha$ defines another pair $x'$ and $\theta$, where $x'$ is the photon path length after the second scattering. The Jacobian of the transformation gives the relation
\begin{equation}
dR'\,sin\alpha\,d\alpha=\frac{R^2}{R'^2}sin\theta|cos\theta|dx'd\theta
\end{equation}
Putting $\Delta S'=4\pi R'^2$, the contribution of photons scattered for the second time at a distance $(R', R'+dR')$ to arrive at an angle $(\theta, \theta+d\theta)$ at the sphere with radius $R$ equals
\begin{eqnarray}
\lefteqn{dn_2^{AB}(\theta,t, x'; R) = e^{-\frac{x'}{\lambda}}\cdot G^{AB}\Big[R'(x',\theta),\alpha(x', \theta)\Big]\cdot }\nonumber \\
&& \cdot 4\pi R^2\cdot 2\pi sin\theta\,|cos\theta| \, \, dx'd\theta dt
\end{eqnarray}
The factor $e^{-x'/\lambda}$ multiplied by $e^{-ct'/\lambda}$ in the expression for $j_1^A$ gives $e^{-ct/\lambda}$, independent of $x'$.
Integration over $x'$ gives the total number of the above photons
\begin{eqnarray}
\lefteqn{dn_2^{AB}(\theta, t; R)=} \\
&&=e^{-\frac{ct}{\lambda}}\int_0^{x'_{max}}G^{AB}_* dx'\cdot 4\pi R^2\cdot 2\pi \,sin\theta\,|cos\theta|d\theta\,dt \nonumber
\end{eqnarray}
where $G_*^{AB}=G^{AB}/ e^{-ct'/\lambda}$.\\
The maximum value of $x'$ results from fixing time $t$. We have that
\begin{equation}
x'_{max}=ct-R'=\frac{R}{2}\cdot \frac{\tau^2-1}{\tau-cos\theta}
\end{equation}
Thus, the flux of the second generation $AB$ equals
\begin{eqnarray}
\lefteqn{j_2^{AB}=\frac{1}{4\pi R^2\cdot 2\pi sin\theta\,|cos\theta|}\frac{d^2n^{AB}_2}{d\theta \,dt}=}\\
&&=e^{-\frac{ct}{\lambda}}\int_0^{x'_{max}}G^{AB}_* dx'\nonumber
\end{eqnarray}
and, integrated over $\theta$ for $\theta<\zeta$, gives $dN_2^{AB}(t;\zeta,R)/dt$.
With the Rayleigh and Mie processes active we must take into account all four cases $A=R$ or $M$ and $B=R$ or $M$. Finally, the flux of the combined second generation photons is a sum of all specific fluxes
\begin{equation}
j_2=j_2^{RR}+j_2^{RM}+j_2^{MR}+j_2^{MM}
\end{equation}
Some of the integrals defined in this Section can be found as analytical functions (Appendix C). It is of some importance when calculating higher generations (see the next Section).

\subsection{The next generations}
Any next generation of the scattered photons can be calculated in the same way as the second one has been found from the previous one (the first). To calculate the flux $j_i(\theta, t; R)$ of the $i-th$ generation, given  $j_{i-1}(\theta, t; R)$ we proceed as before when calculating the second generation from the first one (Eq. 24). The number of photons, incident on $\Delta S'$ at an angle $\theta ', \phi '$ within $d\Omega '$ at time $(t',t'+dt')$ and scattered along $dx$ into $d\Omega (\alpha)$ equals:
\begin{equation}
j_{i-1}(\theta ', t'; R')dt' d\Omega ' \Delta S' |cos\theta'| \frac{dx}{\lambda_i}f_i(\alpha ')d\Omega
\end{equation}
If there are two scattering processes, $R+M$, then 
\begin{equation}
\frac{f_i(\alpha ')}{\lambda_i}=\frac{f^R(\alpha ')}{\lambda_R}+\frac{f^M(\alpha ')}{\lambda_M}
\end{equation}
is the scattering probability by an angle $\alpha '$ by any process per unit distance per unit solid angle.
The rest of the derivation of $j_i(\theta, t ;R)$ is the same as in the previous Section.
However, for each next generation the number of numerical integrations increases, unless values of $j_{i-1}(\theta, t; R) $ are stored as a 3-dimension matrix. Thus, it is convenient to find analytical solutions of the integrals $F(\theta ', \alpha)$ and/or $G(R', \alpha)$, if possible.
\subsection{Results of calculations}
Fig. 3a shows the number of photons arriving at the detector within an angle $\zeta<1^\circ, 3^\circ, 10^\circ$ per unit area per unit $\tau=ct/R$ as  a function of $\epsilon= \tau-1$.
The upper curves refer to the first generation, the lower - to the second one. We also compare here the time distributions obtained for the Rayleigh with those for the isotropic scattering. The distance detector- source equals to one scattering length ($k=R/\lambda=1$).\\
First of all we notice that for short times ($\epsilon\le 0.01$) the first generation dominates over the second one, and (as we can guess) over the higher ones. It can be seen from the formulae for the isotropic scattering (Appendix A) that the ratio $\frac{dN_2^{is}}{d\tau}/\frac{\Delta N_1^{is}}{d\tau}$ for any given time should be proportional to $k=R/\lambda$, so that the importance of the second (and the higher) generation will be bigger for larger $k$.\\
We can also see that the number of photons arriving within an opening angle $\zeta$ reaches the dependence $\sim\zeta^2$ only at later times. This reflects the fact that the initial angular distribution of light is steep and becomes almost flat at times $\tau\ge 1.1$ or so.
When comparing the Rayleigh curves with the isotropic ones one can see that the latter are slightly flatter for shorter times, as might be expected but become parallel to the former for later times. \\
Next, we consider a situation when there are two scattering processes, Rayleigh and Mie with quite different angular distributions $f(\alpha)$ (as discussed before). We adopt

\begin{equation}
k=\frac{R}{\lambda_{tot}}=R ( \frac{1}{\lambda_R}+\frac{1}{\lambda_M} )=1
\end{equation}

and $\lambda_R=\lambda_M$ for simplicity.\\
The result for the first and the second generation depending on time is shown in Fig. 3b. There is now more light at earlier times than in the previous case (Fig. 3a) due to the strong Mie scattering in the forward directions. However, the flux of the first generation decreases about 3 times quicker over the considered time region. Although the ratio of the second to the first generation is practically the same at $\epsilon=10^{-3}$ in both cases $(\le 1\%)$. the importance of the second one is reached at later times when the Mie scattering is present.

\section{Point light source in the atmosphere}
\begin{figure}[th]
  \centering
  \includegraphics[height=3.4in, angle=270]{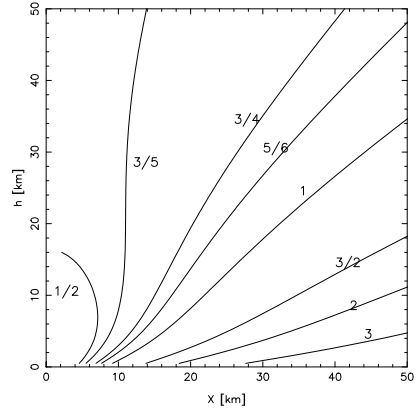}
  \caption{ Vertical cross-section through the atmosphere. Lines correspond to constant values of $k=R/\lambda_{PD}$, shown by numbers, looking from the detector (at $x=0$, $h=0$) to a point on the line, for $\lambda^R=18\,km$ and $\lambda^M=15\,km$ at the ground and the scale heights $H^R=9\, km$ and $H^M=1.2\, km$.}
  \label{fig4}
 \end{figure}
Now we shall study the situation when a point light source flashes in a non-uniform medium, such as the atmosphere. 
We assume that the atmosphere is composed of two sorts of matter, molecules and aerosols, each having its density decreasing with height exponentially with a different scale heights, $H_R$ - for molecules and $H_M$ for aerosols, and 
having the corresponding mean path lengths for scattering at the ground $\lambda_D^R$ and $\lambda_D^M$.
It is not difficult to derive that the effective mean free path for a scattering for light travelling between two arbitrary points $P$ and $S$ equals 
\begin{eqnarray}
\lefteqn{\lambda_{PS}=\Big(\frac{1}{\lambda^R_{PS}}+\frac{1}{\lambda^M_{PS}} \Big)^{-1}=} \\
&&=\frac{h_P-h_S}{\frac{H_R}{\lambda^R_D}\big(e^{-h_S/H_R}-e^{-h_P/H_R}\big)+\frac{H_M}{\lambda^M_D}\big(e^{-h_S/H_M}-e^{h_P/H_M}\big)}
\nonumber
\end{eqnarray}
where $h_P$ and $h_S$ are the heights of points $P$ and $S$ above the level, where the Rayleigh and Mie scattering path lengths are correspondingly $\lambda_D^R$ and $\lambda_D^M$.
If the source (point $P$) is at a distance $R$ from the detector (point $D$ on the ground) then the ratio $k$, of $R$ to the mean free path length along $PD$ equals
\begin{eqnarray}
\lefteqn{k= \frac{R}{\lambda_{PD}}=} \\
=&&\frac{1}{cos\theta_Z}\Big[\frac{H_R}{\lambda^R_D}\big( 1-e^{-h_P/H_R}\big)+\frac{H_M}{\lambda^M_D}\big( 1-e^{-h_P/H_M}\big)\Big] \nonumber
\end{eqnarray}
It can be seen that increasing the distance $R$ to infinity (keeping $\theta_Z$ constant) the ratio $k$ reaches its maximum finite value
\begin{equation}
k_{max}(\theta_Z)= \frac{1}{cos\theta_Z}\Big(\frac{H_R}{\lambda^R_D}+\frac{H_M}{\lambda^M_D}\Big)
\end{equation}
This situation is illustrated in Fig. 4. Here a vertical cross-section of the atmosphere is shown. Detector is at $x=0, h=0$ and lines represent constant values of $k$ corresponding to the straight path from the detector to the point on the line. We have adopted the following values: $\lambda^R_D=18\,km, H_R=9\,km$, $\lambda^M_D=15\,km$ and $ H_M=1.2\,km$. These values describe approximately the atmospheric conditions at the Pierre Auger Observatory \cite{pao}. \\
From Fig. 4 one can also deduce that relevant values of $k$, if light sources are at distances $\sim(10-30)\,km$ (as extensive air showers seen by Auger) are $1/2\le k\le 3/2$. Thus, this will be the region of our interest.
 \begin{figure}[th]
  \centering
  \includegraphics[width=3in]{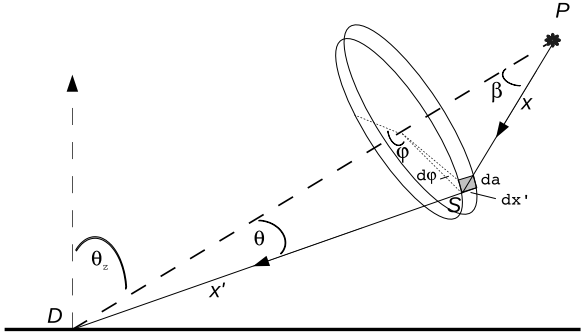}
  \caption{ Geometry of the first generation in the real atmosphere. Light source is at $P$, detector at $D$. Scattering takes place at $S$.}
  \label{fig5}
 \end{figure}
 \begin{figure}[th]
  \centering
  \includegraphics[width=3in]{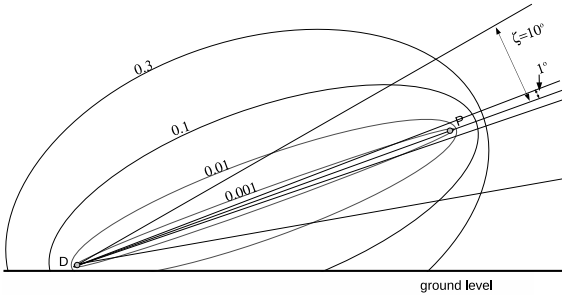}
  \caption{The ellipsoids (their cross-sections are shown) show scattering sites of first generation photons arriving at $D$ after time $\frac{R}{c}(1+\epsilon)$. The corresponding numbers are equal to $\epsilon$. Light source is at $P$, detector at $D$. }
  \label{fig6}
 \end{figure}
\subsection{The first generation}

As the medium is non-uniform, we cannot use the idea of a sphere to be crossed by the scattered photons, as in Section 2. Now their flux will depend on the zenith angle $\theta_Z$ of the source and on the azimuth angle $\phi$ around the direction towards it. Fig. 5 shows a trajectory $PSD$ of a first generation photon scattered at $S$. We want to calculate the angular distribution of the first generation as a function of time, for a fixed $R$ and $\theta_Z$, $\frac{d^2n_1}{d\Omega dt}(\theta,\phi, t;R,\theta_Z)$, crossing a unit area perpendicular to the direction towards the source.\\
We notice that for a fixed arrival direction of photons $(\theta, \phi)$ and time $t$, the scattering point $S$ is uniquely determined. To arrive at the detector at angles $(\theta,\phi)$ within $d\Omega=sin\theta\,d\theta\,d\phi$ after time $(t, t+dt)$, photons have to cross the surface $da$ (shaded in the figure) and be scattered along a path length $dx$ by the angle $\alpha$ determined by Eq.2. The number of such photons equals
\begin{eqnarray}
\lefteqn{dn_1(\theta,\phi,t)=\frac{x'sin\theta\,d\phi\,dx'cos\gamma}{4\pi x^2}
e^{-\frac{x}{\lambda_{PS}}} \cdot} \\
&&\cdot\frac{dx}{\lambda_S}f(\alpha)\Delta \Omega_D e^{-\frac{x'}{\lambda_{SD}}} \nonumber
\end{eqnarray}
where $x=PS$, $x'=SD$, $\gamma$ is the angle between the normal to the surface $da$ and the direction of the incident photons $PS$, $\Delta \Omega_D$ is the solid angle determined by the unit area at $D$ and the scattering point $S$ 
($\Delta \Omega_D=cos\theta/x'^2$). There is no need to calculate $\gamma$ because $dx=x'd\theta/cos\gamma$, so that it cancels out. It can be shown that 
\begin{equation}
\frac{dx'}{x^2}=\frac{2}{\tau^2-2\tau \,cos\theta+1}\,\frac{d\tau}{R}
\end{equation}
Inserting this into (40) we obtain
\begin{eqnarray}
\lefteqn{j_1(\theta,\phi,t)\cdot cos\theta=\frac{d^2n_1}{d\Omega dt}= } \\
&&=\frac{c}{2\pi\lambda_S R^2}\, \frac{f(\alpha)cos\theta}{\tau^2-2\tau \,cos\theta+1}\cdot e^{-(\frac{x}{\lambda_{PS}}+\frac{x'}{\lambda_{SD}})}
\nonumber
\end{eqnarray}
One can see that this formula is practically the same as (7) for the uniform medium, the only difference being in the scattering path lengths depending not only on distances but also on the geometry.\\
The height of the scattering point $S$ necessary to calculate $\lambda_S, \lambda_{PS}$ and $\lambda_{SD}$ equals:
\begin{eqnarray}
\lefteqn{h_s=\Big(ct-R\frac{sin\theta}{sin\alpha}\Big)\cdot} \\
&&\cdot (cos\theta_Z \;cos\theta+sin\theta_Z \;sin\theta \;cos\phi) \nonumber
\end{eqnarray}
We calculate numerically the time distributions of light $dN^{real}_1(t;\zeta)/dt$, arriving at the detector at angles smaller than $\zeta$ for different zenith angles of the source. The results, in the form of the ratio:
\begin{equation}
F_1(\tau; \zeta)=\frac{d N_1^{real}/dt}{d N_1^{uni}/dt}
\end{equation}
are presented in Figs. 7, 8 and 9, where $dN_1^{uni}/dt$ are the distributions obtained in the previous section for a uniform medium.  \\

 \begin{figure}[th]
  \centering
  \includegraphics[height=3in, angle=270]{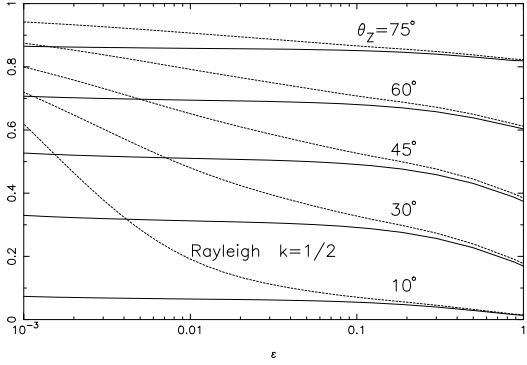}
  \caption{Ratio of the first generation $dN_1^{real}/dt$ in real atmosphere to that in uniform medium $dN_1^{uni}/dt$
as a function of time ($\epsilon=ct/R-1$) for the Rayleigh scattering only for various values of zenith angle of the source. Solid lines - $\zeta=1^\circ$, dashed lines - $\zeta=10^\circ$, $k=R/\lambda^R_{PD}=1/2$.}
  \label{fig7}
 \end{figure}

 \begin{figure}[th]
  \centering
  \includegraphics[height=3in, angle=270]{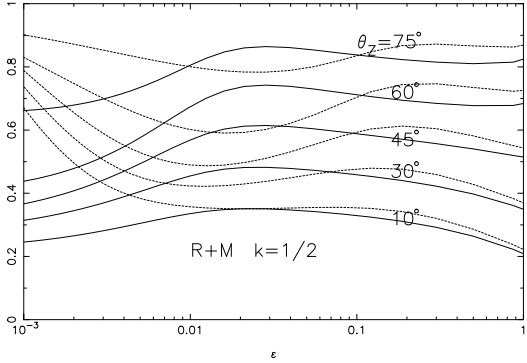}
  \caption{As in Fig. 7 but with Mie included;  $k=R/\lambda_{PD}=1/2$. }
  \label{fig8}
 \end{figure}

 \begin{figure}[th]
  \centering
  \includegraphics[height=3in, angle=270]{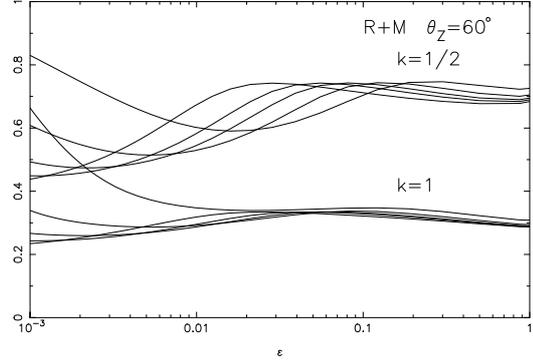}
  \caption{As in Fig. 8 but for two values of $k$. Each group of lines is for $\zeta=1^{\circ},\,2^{\circ},\,3^{\circ},\,5^{\circ}$, and 
$10^{\circ}$ from bottom to top (at $\epsilon=10^{-3}$). }
  \label{fig9}
 \end{figure}
To compare the light flux obtained for the real atmosphere with that for a uniform medium we adopt the same value of $k$ and the same distance from the source to detector for both cases.
To understand the effect of a purely exponential atmosphere we start with considering only the Rayleigh scattering (Fig. 7), neglecting Mie ($\lambda_M=\infty$). Understanding the behaviour of the curves in this figure is easier with the help of Fig. 6. Each ellipse is a cross-section of a rotational ellipsoid with the symmetry axis determined by point $D$ - the detector and point $P$ - the light source. These are the focal points of  all the ellipses. Photons arriving at the detector at $D$ after some fixed time $\tau=1+\epsilon$ (in units of the distance PD) must have been scattered on the surface of an ellipsoid with the eccentricity $e$ equal
\begin{equation}
e=\frac{1}{1+\epsilon}
\end{equation}
The ellipses refer to $\epsilon=10^{-3}, 10^{-2}, 10^{-1}$ and  $3\cdot 10^{-1}$ keeping the right proportions. The detector field of view cuts only a part of the ellipsoid surface where the photons registered after time $1+\epsilon$ must have been scattered. We notice that all ratios $F_1$ in Fig. 7 are smaller than 1 and decrease with time (although those for $\zeta=1^\circ$ are practically constant) and ratios for $\zeta=10^\circ$ are larger than those for $\zeta=1^\circ$. All this becomes clear when inspecting Fig. 6 and the corresponding scattering sites. For example - the constancy of $F_1(\tau; \zeta=1^\circ$) 
for times $\tau-1=\epsilon=10^{-3} \div 10^{-1}$ reflects the fact that the scattering takes place very close to point $P$ during all this time  and starts to move away from it (higher in the atmosphere) only for larger times i.e $\tau\ge 1.3$. Since the scattering path length in the uniform medium is chosen equal to the effective path length in the atmosphere $\lambda_{PD}$ (Eq. 36), we have that $\lambda_P>\lambda_{PD}$ and the scattering probability at $P$ is smaller in the exponential atmosphere.\\
Let us consider now a more realistic atmosphere with both processes, Rayleigh and Mie at work.
Fig. 8 shows time dependence of the ratio $F_1(\tau;\zeta)$ for $\zeta=1^\circ$ and $ 10^\circ$ and zenith angles $\theta_Z=10^\circ\div 75^\circ$. Let us take a closer look at the case $\zeta=1^\circ$ and $\theta_Z=75^\circ$ (upper solid curve). It may seem strange that the ratio $F_1$ increases since in the case of the Rayleigh scattering only it decreases, although very slowly. We have checked that a similar slow decrease takes place if the scattering is only Mie. The behaviour of the curves in this figure would be hard to understand without the presentation of the scattering sites in Fig. 6. It can be seen that for $\epsilon$ smaller than a few $\times 10^{-2}$ the scattering sites are close to the source at $P$. Thus, we may approximate the ratio $F_1$ as follows
\begin{equation}
F_1=\simeq \frac{\frac{\overline f_R}{\lambda_P^R}+\frac{\overline f_M}{\lambda_P^M}}{\frac{\overline f_R}{\lambda_{PD}^R}+\frac{\overline f_M}{\lambda_{PD}^M}}=
\frac{\lambda_{PD}^R}{\lambda_{P}^R}\cdot \frac{1+\frac{\lambda_P^R}{\lambda_P^M}\frac{\overline f_M}{\overline f_R}}{1+\frac{\lambda_{PD}^R}{\lambda_{PD}^M}\frac{\overline f_M}{\overline f_R}}
\end{equation} 
where $\overline f_{R(M)}$  are some effective angular distributions of photons scattered by Rayleigh (Mie) on the cut surface .
As time increases, none of the $\lambda's$ changes much. However, the typical scattering angles increase, what affects much more $\overline f_M$ than $\overline f_R$, so that $\overline f_M/\overline f_R$ decreases. Since
\begin{equation}
\frac{\lambda_{P}^R}{\lambda_{P}^M}<\frac{\lambda_{PD}^R}{\lambda_{PD}^M}
\end{equation}
 the numerator decreases by a smaller factor than the denominator so that the ratio $F_1$ increases. It can be seen from Fig. 6 that for $\epsilon \ge 2\cdot 10^{-2}$ the scattering angles of the registered photons do not change much  (the denominator stays constant) but now $\lambda_P^R$ and $\lambda_P^M$ have to be substituted by $\lambda_S^R$ and $\lambda_S^M$, where $S$ is an effective scattering point with growing height. As $H_M<H_R$, the ratio $\lambda_S^R/\lambda_S^M$ decreases and so does $F_1$. \\
A different behaviour of $F_1(\tau; \zeta=10^\circ)$ can be explained also with the help of Fig. 6. At first $F_1$ decreases (the smaller $\theta_Z$ - the stronger decrease) because the detector field of view cuts out a growing part of the deep atmosphere where the scattering is strong in the real atmosphere. At $\epsilon\ge 0.02$ $F_1$ starts to increase for the same reason as just described in the case $\zeta=1^\circ$. It must finally decrease since the scattering takes places further and further behind the source, where $\lambda^R$ and  $\lambda^M$ are growing in the real atmosphere.\\
In Fig. 9 we show $F_1(\tau; \zeta)$ for $\theta_Z=60^\circ$ for several intermediate values of $\zeta$, and for two values $k=1/2$ and $1$. Note that changing $k=R/\lambda_{PD}$ must result in changing $R$ - the distance to the source. This is the main reason why the curves for $k=1$ are lower than those for $k=1/2$. It can be seen from Fig. 5 that for $\theta_Z=60^\circ$ (the scales on both axes are the same so in the figure the angles are correctly represented) the distance $R$ is much shorter for $k=1/2$ than for $k=1$, implying that the corresponding heights of the source differ considerably. Inspecting Fig. 6 we can estimate that for $\epsilon\simeq 0.01$ and $\zeta=1^\circ$ the curves for $k=1$ should be down with respect to those for $k=1/2$ by a factor
\begin{equation}
\frac{\lambda_{P_1D}}{\lambda_{P_1}}/\frac{\lambda_{P_{1/2}D}}{\lambda_{P_{1/2}}}
\end{equation}
where $P_{1/2}$ and $P_1$ are positions of the source referring to $k=1/2$ and 1 respectively, with the meaning of all $\lambda's$ as defined before.
In our example $\lambda_{P_1D}\simeq 33\,km$, $\lambda_{P_1}\simeq 113\,km$, $\lambda_{P_{1/2}D}\simeq 15.3\,km$, $\lambda_{P_{1/2}}\simeq 25.6\,km$ so that the above factor equals $\simeq 0.49$ whereas the exact ratio from Fig. 9 $\simeq 0.48$.

\subsection{The second generation}
 \begin{figure}[th]
  \centering
  \includegraphics[height=3in, angle=270]{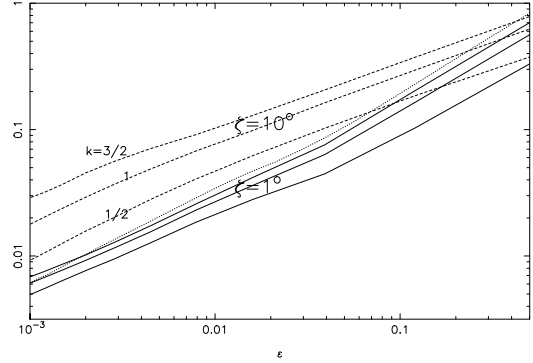}
  \caption{ Ratio of the second $dN_2/dt$ to the first generation $dN_1/dt$ as a function of time ($\epsilon=ct/R-1$) from a flash (at $t=0$) of a point source at zenith angle $\theta_Z=75^\circ$ in the real atmosphere. Fluxes are integrated within $\zeta=1^\circ$ (solid lines) and $\zeta=10^\circ$ (dashed lines). Three curves for each $\zeta$ refer to $k=1/2, 1, 3/2$ (from bottom to top). Dotted line refers to a uniform medium with Rayleigh scattering only, for $k=1$, $\zeta=1^\circ$.}
  \label{fig10}
 \end{figure}
We proceed like in the case of the first generation (Fig. 5), but now point $S$ refers to the second scattering. Photons scattered only once arrive at the surface $da$ from all directions according to $j_1(\theta_1,\phi_1,t_1;x)$, where $t_1=t-x'/c$. Thus, the number of photons $dn_2$ incident on $da$ and scattered for the second time towards the detector (to arrive there within $d\Omega_D(\theta,\phi)$ after time $(t, t+dt)$ equals
\begin{eqnarray}
\lefteqn{dn_2(\theta,\phi,t;x')=\int_{\Omega_1} j_1(\theta_1,\phi_1,t;x)d\Omega_1dt\,da\,cos\gamma\cdot} \nonumber \\
&\cdot&\frac{dl}{\lambda_S}f(\alpha_2)d\Omega_De^{-\frac{x'}{\lambda_{SD}}} 
\end{eqnarray}
where the integration has to be done over full solid angle $(0< \phi_1< 2\pi, 0\le \theta_1\le\pi)$, $dl$ is the path length for the second scattering to occur ($cos\gamma \, dl=x'\,d\theta$) and $\alpha_2$ is the angle of the second scattering.
Now the pair of variables, $\theta$  and $t$, does not determine uniquely the positions of the second scattering, since the times $t_1$ elapsed from photon  emission to their arrival at $S$ (or strictly speaking, at $da$) have some distribution. However, $t_1$ can not be smaller than $x/c$, thus $x'_{max}=ct-x$. Expressing $x'_{max}$ as a function of $\theta$, $t$ and $R$ only we obtain  
\begin{equation}
x'_{max}=\frac{(ct)^2-R^2}{2(ct-R\,cos\theta)}
\end{equation}
and the distribution of the second generation  equals
\begin{eqnarray}
\lefteqn{\frac{d^2n_2(\theta,\phi,t)}{d\Omega\,dt}=\int_0^{x'_{max}}dx'\int_0^{2\pi}d\phi\cdot} \\
&&\cdot\int_0^{\pi}d\theta_1\,j_1\,sin\theta_1\cdot\frac{f(\alpha_2)}{\lambda_S}
e^{-\frac{x'}{\lambda_{SD}}}cos\theta \nonumber
\end{eqnarray}

Finally, the number of photons within an angle $\zeta$ equals
\begin{equation}
\frac{dN_2(t;\zeta)}{dt}=\int_0^{2\pi}\int_0^{\pi}
\frac{d^2n_2}{d\Omega\,dt}sin\theta\,d\theta\,d\phi
\end{equation}
The above integrals have been calculated numerically.\\
The ratio $\frac{dN_2}{dt}/\frac{d N_1}{dt}$ as a function of $\tau-1$ in the real atmosphere, where the source is at $\theta_Z=60^\circ$ at the distance $R=34\,km $ corresponding to $k=1$, within two opening angles of the detector $\zeta=1^\circ$ and $10^\circ$, is presented in Fig. 10. For comparison we have also drawn there the same
ratio for the uniform medium, for $k=1$ with Rayleigh scattering active only (following from Fig. 3a for $\zeta=1^\circ$). First we see that the contribution of the second generation increases with time as it should be expected. Next, it is considerably larger for the uniform medium than for the real atmosphere, which may not be so obvious at first sight. In fact, at $\epsilon=10^{-2}\div 10^{-1}$ the curve for the uniform medium is $\sim 2.3$ times higher than that for the real atmosphere. This factor equals ( for simplicity we neglect the operators $\frac{d}{dt}$)
\begin{eqnarray}
\lefteqn{\eta=\frac{N_2^{uni}(R)}{N_1^{uni}(R)}/\frac{N_2^{real}(R+M)}{N_1^{real}(R+M)}=}\\
&&=\frac{N_1^{real}(R+M)}{N_1^{uni}(R)}/\frac{N_2^{real}(R+M)}{N_2^{uni}(R)}\nonumber
\end{eqnarray}
The numerator $\frac{N_1^{real}}{N_1^{uni}} \sim \frac{\lambda_{P_1D}}{\lambda_{P_1}}= 0.29$ as explained in the previous paragraph. The denominator refers to photons scattered exactly two times. For small $\epsilon$ the scatterings must take place close to the source or along the field of view ($\zeta=1^\circ$ in our example). Thus we should have that 
\begin{equation}
\frac{N_2^{real}(R+M)}{N_2^{uni}(R)}\ge \Big(\frac{\lambda_{P_1D}}{\lambda_{P_1}}\Big)^2=0.084
\end{equation}  
Our calculations show that
$\frac{N_2^{real}}{N_2^{uni}}\simeq 0.13$ for $\epsilon=$ few $\times 10^{-2}$ so that $\eta\sim\frac{0.29}{0.13}\simeq 2.2$, in agreement with the exact calculations. 

\section{A moving point source - a shower}
So far we have been interested in the light signals from a flash of a stationary point source in a detector at a given distance, within a given field of view, as a function of time.\\
Now, we shall consider a moving point emitting light on its way through the atmosphere. This is a reasonable model of a distant cosmic ray shower exciting atmosphere what results in emitting isotropic fluorescence light. 
Moreover, the Cherenkov light produced by $\sim 1/3$ of shower electrons \cite{espec}, propagating alongside the particles will be scattered to the sides by the Rayleigh and Mie processes in an anisotropic way. Thus, in principle, a shower is a moving source, emitting light anisotropically. Our aim is to find an instantaneous optical image of the shower, produced by the scattered light at \emph{the same time} as the shower image in the direct (not scattered) light. \\
The image obtained for a small integration time $\Delta t$ is a radially symmetric light spot, corresponding (roughly) to the emission distribution at the shower lateral cross-section \cite{somm}.
It is produced mainly by the direct light. But at the same time as the 
direct light, there arrive also photons produced by shower particles at an earlier stage of shower evolution and scattered in the atmosphere. As shower parameters follow straightforwardly from the 
amount of the direct light, it is important to calculate this effect and to 
take it into account when deriving the flux of the direct light.\\
Let us assume that we know the geometry and the timing of the shower. If 
the source is at point $Q$ at time $t$ and emits $C$ photons per unit length, the telescope will record the following 
number of the direct photons at time $t+R/c$, in the time bin $\Delta t$:

\begin{equation}
\Delta n_0=C \cdot\frac{c\Delta t}{1-cos\delta}f_P(\delta)\frac{A}{R^2}e^{-\frac{R}{\lambda_{QD}}}
\end{equation}

where $\delta$ is the angle between the shower direction and the direction 
from point $Q$ to the detector $D$, $\lambda_{QD}$ is the effective mean free 
path for attenuation along the distance $R$, $A$ is the 
diaphragm area determining the collection solid angle and $f_P(\delta)$ is the angular distribution 
of the emitted light.\\
In the case of isotropic fluorescence light $f_P(\delta)=1/4\pi$ and $C$ equals to the number of photons produced 
per unit length ($\sim 4\,m^{-1}$, per one electron). In the case of the
scattered Cherenkov photons, or those from a laser, $f_P(\delta)=\frac{3}{16\pi}(1+cos^2\delta)$ for the Rayleigh scattering
(or a corresponding distribution for Mie) and 
$C=N/\lambda_Q$, where $\lambda_Q$ is the mean free path (in length units) for the 
process in consideration at point $Q$ and $N$ is the number of Cherenkov photons at the observed shower level propagating (approximately) in the direction of the shower. The number of the scattered Cherenkov photons to the fluorescence ones is typically $10-15\%$, and a lot of the former are scattered by Rayleigh, what is not far from isotropic. Thus, with a good approximation one can treat a shower as a moving isotropic light source. This is what has been adopted in this paper: $f_P(\delta)=\frac{1}{4\pi}$.\\
Simultaneously with the direct light some photons produced earlier (above point $Q$) will also 
arrive at the telescope. However, as we already mentioned, they must have 
been scattered on their way to the telescope.
Our earlier calculations \cite{giller} showed that if the distance to the shower is 
not very much longer than the mean scattering path, then the main 
contribution to the scattered light is due to the photons scattered only 
once (the first generation). Here, we will calculate the fraction of 
the total light consisting of the first and the second generations.
\subsection{The first generation} 
Let us consider photons produced at point $P$, along a shower path element 
$dl$ (Fig. 11). The first generation photons arriving at the detector
$D$ at the same time as the direct photons produced at $Q$ must have been scattered
on the surface of the rotational ellipsoid with the focal points at $D$ and $P$, with the $DP$ line being the 
rotational symmetry axis of the ellipsoid. \\
As we have already mentioned, the main interest of the shower experiments 
is to measure the direct light. Thus, we are interested in the light 
arriving from the direction of point $Q$ and the region around it. In 
practice (as it is in The Pierre Augere Experiment) the light signal 
arriving simultaneously with the direct light from point $Q$ is measured 
within a certain angle $\zeta$ around the line $DQ$. The cone with the 
opening angle $\zeta$ cuts out on the surface of the ellipsoid a region of 
our interest (the shaded surface). It is at this part of the 
ellipsoid that the photons produced at $P$ have to be scattered to 
arrive simultaneously with the direct light produced at $Q$, within an 
angle smaller than $\zeta$ with respect to the latter. Thus, to find the 
contribution of the first generation to the direct light one has to 
integrate the number of the scattered photons over the cutout surface, and 
then integrate the result over the distance $QP=l$ (the upper part of the 
shower).\\
 \begin{figure}[th]
  \centering
  \includegraphics[height=3in]{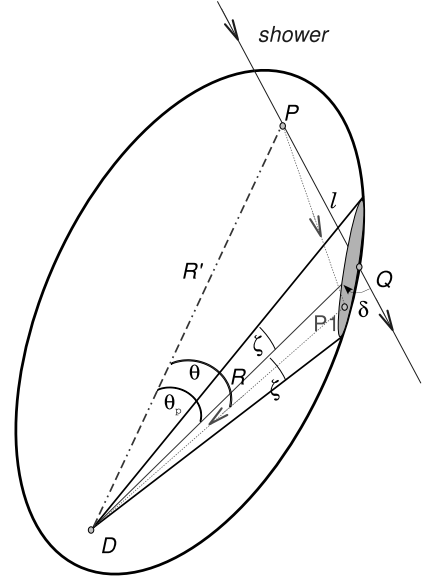}
  \caption{ First generation from a shower. The shaded surface shows sites of scattering of photons produced at $P$, scattered once and arriving at detector $D$ within its field of view simultaneously with direct photons produced at $Q$. $PQ+QD=PP_1+P_1D$.}
  \label{fig11}
 \end{figure}

\begin{figure}
  \centering
  \includegraphics[width=3in]{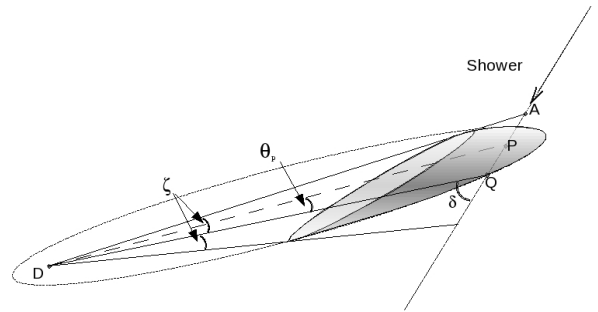}
  \caption{Fluorescence light produced at point $P$ and scattered on the shaded surface (inside field of view of detector $D$) of the ellipsoid arrives at $D$ simultaneously with the direct light produced at $Q$.  }
  \label{fig12}
\end{figure}
The proportions in Fig. 11 have not been 
preserved. For example the actual distance from point $P$ to the detector 
at $D$ is much longer than $l=QP$ (although in 
principle the integration over $l$ should go far up the shower, the 
distances contributing to the total scattered light are rather close to 
point $Q$). Also the angle $\zeta$ does 
not need to be larger than a few degrees, but on the figure it is much larger 
for the sake of clarity.\\
Let us first find the contribution to the first generation from photons produced by the shower at point $P$ along a path length element $dl$, assuming that $P$ lies outside the detector field of view.
\paragraph{\emph{a)} $\theta_P>\zeta$ \emph{(Fig. 11)}}
\begin{eqnarray}
\lefteqn{d\Big(\frac{d n_1}{dt}\Big)=\frac{cCdl}{2\pi R'^2} \cdot\int_{\theta_P-\zeta}^{\theta_P+\zeta}d\theta
\Bigg[\frac{f(\alpha)\ sin\theta \ cos\theta}{\tau'^2-2\tau'cos\theta+1} \cdot  } \nonumber \\
&&\cdot\int_0^{\phi_{max}(\theta)} \frac{e^{-(\frac{PP_1}{\lambda_{PP_1}}+\frac{P_1D}{\lambda_{P_1D}})}}{\lambda_{P_1}} d\phi\Bigg]
\end{eqnarray}
where $\tau'=\frac{l+R}{R'}$, $tg(\alpha/2)=\frac{\tau'-cos\theta}{sin\theta}$, $cos\phi_{max}=\frac{cos\zeta-cos\theta_P \ cos\theta}{sin\theta_P \ sin\theta} $, $\theta$ is the arrival angle (with respect to the direction to the source at $P$),
and $R'=PD=\sqrt{R^2+l^2-2Rlcos\delta}$.\\
The azimuth angle $\phi$ is measured in the plane perpendicular to the axis $PD$ and $\phi=0$ refer to points on the shower-detector plane (Fig. 11 plane). The length $PP_1$ can be found from the triangle $PP_1D$: $PP_1=R'sin\theta/sin\alpha$ and $P_1D=l+R-PP_1$.
\paragraph{\emph{b)} $\theta_P<\zeta$} (Fig. 12).\\
It is clear from the figure that now the integration limits of $\theta$ and $\phi$ are different so that one has to add to the integral in case a) the following term

\begin{eqnarray}
\lefteqn{\frac{cCdl}{2\pi R'^2}\int_{0}^{\zeta-\theta_P}d\theta\Bigg[\frac{f(\alpha)\ sin\theta \ cos\theta}{\tau'^2-2\tau'cos\theta+1} \cdot}\\
&&\quad\quad\quad\cdot 2 \int_0^{\pi} \frac{e^{-(\frac{PP_1}{\lambda_{PP_1}}+\frac{P_1D}{\lambda_{P_1D}})}}{\lambda_{P_1}}       d\phi        \Bigg]\nonumber
\end{eqnarray}
and the lower limit of $\theta$ in the term a) changes sign, so that $\theta_P-\zeta \to \zeta-\theta_P$.
To find the total flux of the first generation arriving at the same time as the direct light from point $Q$ one has to integrate over $l$ the contribution from case b) from 0 to $l_{max}=R sin\zeta/sin(\delta-\zeta)$, (where $\delta\ge \zeta$), and add to it the contribution from a) integrated from $l_{max}$ to $\infty$. (It turns out that 
contribution from points with $l\simeq 0.1 R$ are negligible). \\
If there are two scattering processes the first generation means the sum of the number of photons scattered by Rayleigh and those by Mie and then $\frac{f(\alpha)}{\lambda_{P_1}}=\frac{f^R(\alpha)}{\lambda_{P_1}^R}+\frac{f^M(\alpha)}{\lambda_{P_1}^M}$, and in the attenuation exponent $\lambda_{PP_1}$ and $\lambda_{P_1D}$ are the effective mean free paths for both processes.
\subsection{Analytical calculation of $\,\frac{d n_1^R(\zeta)}{dt}$}
By making some approximations we have found an analytical solution of the shower image in the first generation light scattered by the Rayleigh process.\\
We shall treat separately light produced within the detector field of view $\frac{dn^R_{in}}{dt}$ and that outside it, $\frac{dn^R_{out}}{dt}$, starting with the former (Fig. 12). As the scattering points lie close to point $Q$ we assume that $\lambda_{P_1}=\lambda_Q$. For the same reason we assume that the exponential factor describing light attenuation equals $e^{-\frac{R}{\lambda_{QD}}}$. Our main assumption, however, consists in integrating over the symmetric (with respect to axis $PD$) part of the ellipsoid so that we could solve the integrals over $\phi, \theta$ and finally $l$. Expressing scattering angle $\alpha$ as a function of $\theta$ and $\tau'$ (Eq. 2) we obtain 
that the contribution from a shower path element $dl$ equals
\begin{eqnarray}
\lefteqn{d\Big(\frac{dn_{in}^R}{dt}\Big)\simeq\frac{Bdl}{R'^2}2\pi\frac{3}{8\pi}\cdot } \\
&&\cdot \int_0^{\theta_{max}}\big(1-\frac{2sin^2\theta}{y'}+\frac{2sin^4\theta}{y'^2}\big)\frac{sin\theta\ cos\theta\ d\theta}{y'}
\nonumber
\end{eqnarray}
where $y'=\tau'^2-2\tau'cos\theta+1$, $B=\frac{cC}{2\pi\lambda_Q}e^{-\frac{R}{\lambda_{QD}}}$.\\
We adopt $\theta_{max}=\zeta$ being the mean of the limiting values of $\theta$ in the exact integration.\\
In the air shower experiments it is the direct light which should be measured so that the viewing angle $\zeta$ is small (a few degrees) and so are angles $\theta$. This leads to $\tau'-1\equiv\epsilon\ll1$.
Taking all this into account the result of the integration over $\theta$ is

\begin{eqnarray}
\lefteqn{ d\Big(\frac{dn_{in}^R}{dt}\Big)\simeq\frac{cCe^{-\frac{R}{\lambda_{QD}}}}{\lambda_Q} \frac{3}{8\pi}\cdot } \\
&&\cdot\frac{1}{2}\Bigg[ln\frac{x^2+1}{x^2}-\frac{1}{(x^2+1)^2}\Bigg]\frac{dl}{R'^2}\nonumber
\end{eqnarray}
where $x=\epsilon/\zeta$. From geometrical considerations it can be derived that
\begin{eqnarray}
\lefteqn{\frac{dl}{R'^2}=\frac{1}{R}\Bigg[\frac{1}{1-cos\delta}-\frac{\epsilon\,cos\delta}{(1-cos\delta)^2}\Bigg]d\tau'\simeq} \\
&&\simeq\frac{d\tau'}{R(1-cos\delta)} =\frac{\zeta dx}{R(1-cos\delta)} \nonumber
\end{eqnarray}
where we have dropped the second term $\sim \epsilon$.\\
To obtain $\frac{dn_{in}^R}{dt}$ it remains to integrate (59) over $x$ (including (60) from $0$ to $x_{max}$, corresponding to point $P$ lying  on the edge of the viewing cone ($\theta_P=\zeta$). Again, assuming that $\zeta\ll 1$, it can be derived that $x_{max}\simeq tg\frac{\delta-\zeta}{2}$ (for $\delta>\zeta$).
The result of the integration is the following
\begin{eqnarray}
\lefteqn{\frac{dn_{in}^R}{dt}\simeq \frac{cCe^{-\frac{R}{\lambda_{QD}}}\zeta}{4\pi R\lambda_Q (1-cos\delta)}\cdot}\\
&& \cdot\frac{3}{16}\Big(3\delta'-sin\delta'-8tg\frac{\delta'}{2}\cdot ln\,sin\frac{\delta'}{2}\Big)\nonumber
\end{eqnarray}
where $\delta'=\delta-\zeta$. \\
\begin{figure}[th]
  \centering
  \includegraphics[height=3in]{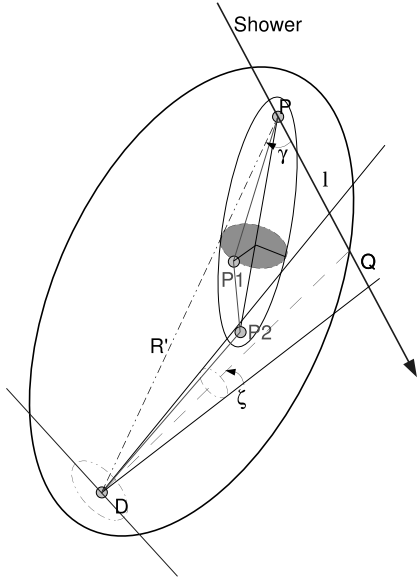}
  \caption{Second generation from a shower. Photon, emitted at $P$ must be scattered for the second time (point $P_2$) within detector $D$ field of view. First scattering ($P_1$) must occur on the surface of the ellipsoid with focal points at $P$ and $P_2$. Direct light from $Q$ arrives at $D$ at the same time: $PQ+QD=PP_1+P_1P_2+P_2D$.}
  \label{fig13}
\end{figure}
To calculate the contribution from the outer part of the shower i.e. outside the field of view we keep the previous approximate assumptions. However, since now the scattering surface looks differently (Fig. 11) we make another simplifications. We assume that all scattering angles equal $\delta$ so that $f(\alpha)=f(\delta)=const$. Next, as $\zeta$ is small we assume that all $\theta=\theta_P$. Integration over $\Omega_D$, the solid angle of the field of view, is then reduced to multiplying the integrand for $\theta=\theta_P$ by $\Omega_D=\pi \zeta^2$. Thus we have
\begin{eqnarray}
\lefteqn{d\Big(\frac{dn_{out}^R}{dt}\Big)= \frac{cCe^{-\frac{R}{\lambda_{QD}}}dl}{2\pi \lambda_Q R'^2}\cdot 
\int_{\Omega_D}\frac{f(\alpha)d\Omega(\theta,\phi)\,cos\theta}{\tau'^2-2\tau'cos\theta+1}\simeq}\nonumber \\
&&\simeq \frac{cCe^{-\frac{R}{\lambda_{QD}}}}{2\pi}\frac{f^R(\delta)}{\lambda_Q}
\frac{\pi\zeta^2cos\theta_P}{\tau'^2-2\tau'cos\theta+1}\frac{dl}{R'^2} 
\end{eqnarray}
It can be derived that the denominator in the above expression equals
\begin{equation}
R'^2(\tau'^2-2\tau'cos\theta+1)\simeq(1+sin^2\delta-cos\delta)l^2
\end{equation}
Assuming\footnote{Taking $cos\theta_P\simeq 1-\frac{1}{2}sin\delta(\frac{l}{R})^2$ needs to allow for light attenuation along $l$ in the integration over $l$. The correction to $dn_{out}/dt$ is a few percent, being negligible for the total $dn^R_1/dt$.} that $cos\theta_P=1$, we obtain 
\begin{eqnarray}
\lefteqn{d\Big(\frac{dn_{out}^R}{dt}\Big)\simeq\frac{cCe^{-\frac{R}{\lambda_{QD}}}}{2\pi}\frac{f^R(\delta)}{\lambda_Q} \cdot} \nonumber\\
&&\cdot\frac{\pi\zeta^2}{1+sin^2\delta-cos\delta}\int_{l_{min}}^\infty\frac{dl}{l^2}
\end{eqnarray}
Since $l_{min}\simeq\frac{R\zeta}{sin(\delta-\zeta)}$ we have
\begin{equation}
\frac{dn_{out}^R}{dt} \simeq \frac{cCe^{-\frac{R}{\lambda_{QD}}}\zeta}{2\pi R\lambda_Q}\frac{3}{16}\frac{(1+cos^2\delta)sin(\delta-\zeta)}{1+sin^2\delta-cos\delta}
\end{equation}
and
\begin{equation}
\frac{d n_1^R}{dt}= \frac{dn_{in}^R}{dt}+\frac{dn_{out}^R}{dt}
\end{equation}
Expressing the number of photons $\Delta n_1^R=\frac{d n_1^R}{dt}\Delta t$ as the ratio to that of the direct light (Eq. 54 for $f(\delta)=1/4\pi$)
we finally obtain
\begin{eqnarray}
\frac{\Delta n_1^R}{\Delta n_0}&\simeq &\frac{3}{16}\, \frac{R}{\lambda_Q}\zeta \Bigg[3\delta'-sin\delta' -8tg\frac{\delta'}{2}\cdot ln(sin\frac{\delta'}{2})\nonumber \\
&+&2\frac{sin\delta'(1+cos^2\delta)(1-cos\delta)}{1+sin^2\delta-cos\delta}\Bigg]
\end{eqnarray}
where $\delta'=\delta-\zeta$ (in radians). Terms $\propto \zeta^2$ and of higher order have been neglected. We can see that, with the approximations adopted, the ratio $\Delta n_1^R/\Delta n_0$ is proportional to $R\zeta/\lambda_Q$ reflecting the fact (as we shall see below) that most scatterings take place close to point $Q$ since $R\zeta$ is proportional to the shower path segment seen by the detector. For $\zeta\ll \delta, \ \delta'\simeq\delta$ and the dependence on $\delta$ separates from that on other parameters. It becomes also obvious that it must be different for different angular distributions $f(\alpha)$, of photons at scattering. As $f^R(\alpha)\propto 1+cos^2\alpha$ is quite different from $f^M(\alpha)$ which is peaked in forward directions, we expect another dependence on $\delta$ for the latter.\\
While deriving the analytical formula (66) we have also assumed that the length of the shower segment cut out by the detector field of view is small when compared  to the distance $R$. For $\zeta\ll 1$ it is fulfilled for almost all $\delta$, apart from the case when $\delta$ is close to 0 or $\pi$. We have calculated analytically the ratio 
$\Delta n_1^R/\Delta n_0$ for $\delta=\pi-\gamma$ for $\gamma\to 0$. However, as the derivation is lengthy and (as it will be seen later) agrees with the exact, numerical results only at $\delta=\pi$, we present here only the final result:
\begin{equation}
\frac{\Delta n_1^R}{\Delta n_0}_{\overrightarrow{\delta\to\pi}}\, \frac{9}{16}k\,\pi\,\zeta\big(1-\frac{7}{3\pi}\zeta\big)
\end{equation}
These values are marked as stars (for $\delta=\pi$) in Fig. 15.\\
 \begin{figure*}[th]
  \centering
  \includegraphics[width=4in, angle=270 ]{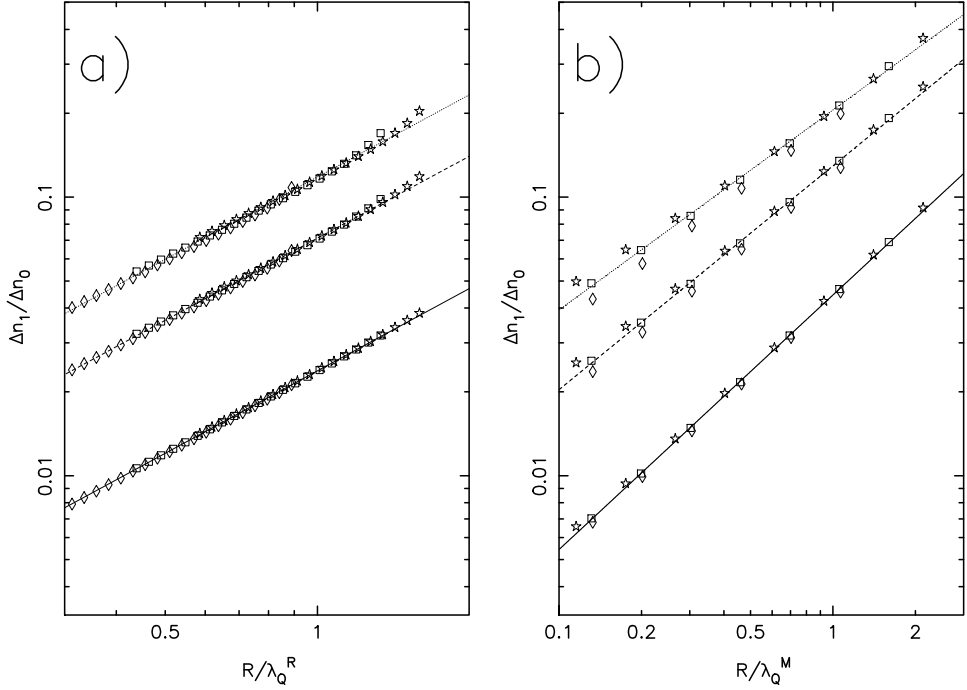}
  \caption{Ratio of the first generation to the direct light as a function of distance $R$ to point $Q$ on the shower in units of the scattering length $\lambda_Q$ at that point. \textbf{a).} Rayleigh only \textbf{b).} Mie only. Curves correspond to $\zeta=1^\circ, 3^\circ, 5^\circ$ (from bottom to top) and $\delta=90^\circ$. Various point shapes refer to different distances $R$=12, 24, 32 $km$. Lines are power law fits.   }
  \label{fig14}
 \end{figure*}
 \begin{figure*}[th]
  \centering
  \includegraphics[width=4in, angle=270]{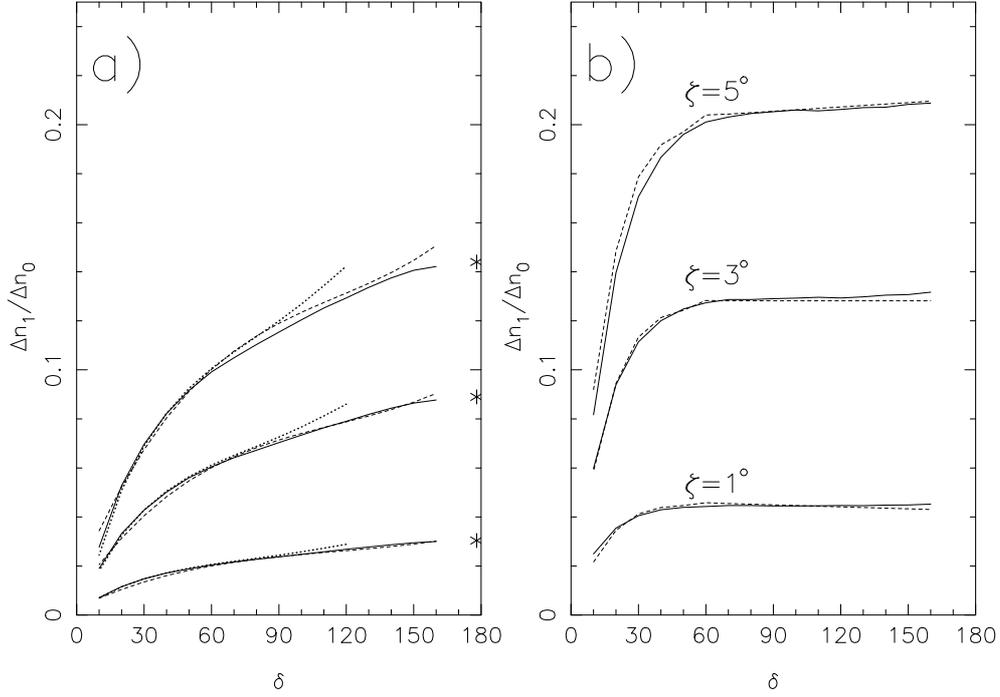}
  \caption{ Ratio of the first generation to the direct light as function of angle $\delta$ between shower and line of sight. \textbf{a).} Rayleigh only \textbf{b).} Mie only. Solid lines - exact numerical calculations, dotted lines and stars in a) - analytical (approximate) calculations, dashed lines - our parametrisation of the numerical curves. All curves are for $k_Q=R/\lambda_Q=1$.
}
  \label{fig15}
 \end{figure*}
We have not tried to calculate $\Delta n_1^M/dt$ because the function $f^M(\alpha)$ is not well known, depending on the sizes of the aerosol particles and, anyway, our analytical approach is approximate. Thus, it seems better to calculate it exactly numerically and, being led by our analytical solutions for Rayleigh, find an appropriate parametrisation of the numerical results.
\subsection{Numerical (exact) calculations of $\Delta n_1/\Delta n_0$ and $\Delta n_2/\Delta n_0$}
The ratio of the first generation to the direct light arriving simultaneously is obtained by numerical integration of point contributions over  $l$  (shower track above point $Q$ ) and dividing the result by $\Delta n_0$. Fig. 14 shows the ratio for the Rayleigh a) and Mie b) scattering as a function of  $k_Q = R/\lambda_Q$ for 
$\zeta = 1^\circ,  3^\circ$  and $5^\circ$ and for several different distances $R$. It can be seen that for the Rayleigh case the ratio is proportional to $k_Q$  and $\zeta$, as it has been derived analytically. For Mie the dependence follows a power law, with the  indices depending slightly on $\zeta$. However, in each case there is practically no dependence on the distance $R$ itself. 
These results are for  $\delta= 90^\circ$.\\
The dependence on $\delta$ is shown in Fig. 15a,b for  $k_Q  =$ 1 and  $\zeta = 1^\circ, 3^\circ$  and $5^\circ$. One can see the difference of the behaviour of this dependence between Rayleigh and Mie.
In the figure a comparison of our analytical calculations with the exact ones is also shown.  The biggest difference reaches some $10\%$  at large $\delta$. There is no normalisation there. We find this agreement quite satisfactory.\\
Nevertheless, as the agreement is not perfect we have parametrised the exact numerical results. For the Rayleigh scattering the ratio $\Delta n^R_1 /\Delta n_0$  can be expressed as 
\begin{equation}
\frac{\Delta n_1^R}{\Delta n_0}=0.024\cdot k_Q \cdot\zeta \,g^R(x)
\end{equation}
where  $x=\delta/100^\circ$, $\zeta$ is in degrees, and 
\begin{equation}
g^R(x)=0.112+1.86x-1.33x^2+0.383x^3
\end{equation}
For Mie the factorisation is not as complete as for Rayleigh, and our fit to the numerical results is the following:
\begin{eqnarray}
\lefteqn{\frac{\Delta n_1^M}{\Delta n_0}=} \\
&&=0.096\cdot\big(\frac{k_Q}{0.7}\big)^{0.93-0.04\zeta}\cdot\big(\frac{\zeta}
{3^\circ}\big)^{0.93+0.05\frac{\delta}{90^\circ}}\cdot g^M(x) \nonumber 
\end{eqnarray}
\begin{eqnarray}
x&\le& 0.6 \quad g^M(x)=x(7.76x^2-11.25x+5.625) \nonumber \\
x &>& 0.6 \quad g^M(x)=1
\end{eqnarray}
These fits are also shown in Fig. 14 and 15. It is seen that they do not deviate from the exact values by more than a few percent.\\
The second generation has been calculated numerically. Fig. 13 should be helpful for understanding the calculation. The thin solid line shows a path of a photon produced at $P$, scattered at points $P_1$ and $P_2$ and arriving at $D$.
The point $P_2$ of the second scattering must lie somewhere inside 
the viewing cone. Fixing it at a distance  $r_3$  from the detector    
corresponds to fixing the value of the remaining photon path length $PP_1P_2$. We have that 
                                                                                                                              
\begin{equation}
PP_1+P_1P_2=R+l-r_3
\end{equation}
where $l=PQ$ and $R=DQ$.\\
In order to assure that the photons scattered twice arrive at $D$ at the same time as the direct light  emitted at 
$Q$, point $P_1$  must lie anywhere on the surface of the ellipsoid with focal points at $P$ and $P_2$. 
Thus, to calculate the number of photons produced at $P$ and scattered twice  one has to integrate 
the contributions like in (49) over the position 
of the first scattering (the whole surface of the ellipsoid) and then  integrate the result over the volume of the 
viewing cone. Finally, the integration has to be performed over the distance $l  = PQ$.\\
 \begin{figure*}[th]
  \centering
  \includegraphics[width=4in, angle=270]{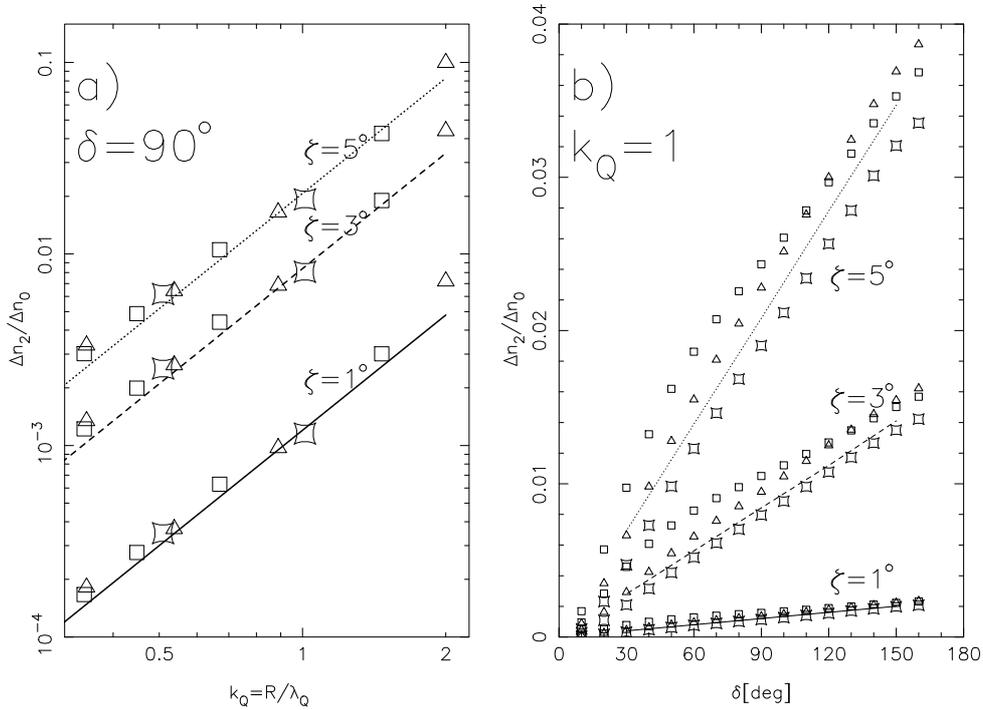}
  \caption{ \textbf{a).} Ratio of the second generation to the direct light as function of $k_Q$.
\textbf{b).} Ratio of the second generation to the direct light as function of angle $\delta$ between shower and line of sight. Points - results of numerical calculations, lines - our parametrisation. Different point signs refer to different distances ($R$ = 12, 24, 32 $km$); $k_Q=R/\lambda_Q=1$.}
  \label{fig16}
 \end{figure*}
The ratio $\Delta n_2/\Delta n_0$, including Rayleigh and Mie scatterings as a function of $k_Q=R/\lambda_Q$ for $\delta=90^\circ$ and as a function of $\delta$ for $k_Q=1$ is presented in Fig. 16. Various point signs refer to different distances $R$. As in the case of the first generation the ratios depend on $R/\lambda_Q$ rather than $R$ itself. However, now they are proportional to the second power of it. We have parametrise the obtained numerical results as follows
\begin{equation}
\frac{\Delta n_2}{\Delta n_0}=1.2\cdot 10^{-3}\, k_Q^2\,\zeta^{1.77}(deg)\cdot\frac{\delta(deg)}{90}
\end{equation} 

\section{Discussion of shower results}
\subsection{Dependence of $\frac{\Delta n_1+\Delta n_2}{\Delta n_0}$ on height}
As we see the numbers of the scattered photons, $\Delta n_1$ and $\Delta n_2$, depend on the angle $\zeta$ within which they arrive at the detector. This angle has to be chosen in such a way as to encompass the total direct signal from the shower. It is obvious that for closer showers $\zeta$ has to be  larger and for those more distant - smaller, so that $R\zeta$ - meaning the lateral spread of the direct light - should remain constant. Our approximations show that in this case ($R\zeta=const$) $n_1^R/n_0$ depends only on height (by $\lambda_Q$) and $\delta$. The power indices in the dependence of $\Delta n_1/\Delta n_0$ on $R$ and $\zeta$ are close to $1$, whereas $\Delta n_2/\Delta n_0\sim (R\cdot\zeta)^2$ (roughly), so that their dependence on the distance $R$ itself should be weak. Fig. 17 presents the contributions of the individual components of the scattered light as a function of height, for $R\zeta=30\,km\cdot deg$, $\delta=90^\circ$ and two (quite different) values of $R$: 12 and 36 $km$. The dependence of the sum $\frac{\Delta n_1+\Delta n_2}{\Delta n_0}$ on $R$ is very weak, indeed, the more so as the ratios $\Delta n_1^M/\Delta n_0$ and $\Delta n_2/\Delta n_0$ change in opposite directions. The chosen value $R\zeta=30\,km\,deg$ corresponds to a shower lateral radius of $\sim 500\,m$, equal to $\sim5\,r_M$ (Moli\`ere radii). This may seem too large to contain the total fluorescence signal. However, deep in the atmosphere it is the shower Cherenkov light what determines its lateral dimensions and it goes up to $5\,r_M$ \cite{latch}.\\
As the typical viewing angles $\delta$ are not far from $90^\circ$ one can draw a conclusion that the maximum contribution of the MS light is $\sim 14\%$ for $R\zeta=30\,km \,deg$ and the adopted atmospheric parameters.
\subsection{Influence on a reconstruction of shower parameters, $E_0$ and $X_{max}$ }
 \begin{figure}[th]
  \centering
  \includegraphics[height=3in, angle=270]{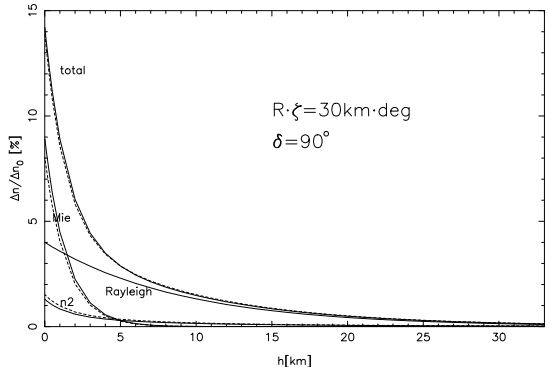}
  \caption{ Ratio of the first two generations to the direct light as a function of height of the observed shower. Solid lines - $R=12\,km$, dashed lines $R=36\,km$. Curves denoted by "Rayleigh" and "Mie" refer to the first generation, by "n2" - to the second one. If a constant lateral dimension is chosen ($R\,\zeta=const$, $\delta=90^\circ$) the ratios do not (practically) depend on shower distance $R$.
}
  \label{fig17}
 \end{figure}
Let us estimate how not allowing for the MS effect would affect a reconstructed  value of a shower primary energy $E_0$. The total signal would then be treated as the direct light, therefore the primary energy would be overestimated. To estimate how much it would be let us consider a typical (for Auger) shower with $E_0=10^{19}$ eV, falling to the ground 25\,$km$ away from the light detector, at $\theta_Z=40^{\circ}$ and azimuthal angle $90^{\circ}$ (angle between the direction from the detector to the shower core and the projection of the shower axis on the ground). If the primary is a proton the mean depth of its maximum equals $X_{max}\simeq 790\,g\,cm^{-2}$, corresponding to the height above Auger level ($\sim 860\,g\,cm^{-2}$) $h_{max}\simeq 3.2\,km$. From Fig. 17 we see that the contribution of the scattered light is $\sim4.5\%$ if  $R\zeta=30\,km\,deg$, what corresponds to $\zeta=1.2^{\circ}$.
However, in the Auger telescopes the individual pixel has a diameter of $1.5^{\circ}$, so that it would be necessary to choose also the neighbouring pixels around the one with the stronger signal to collect the total direct light. Thus, we choose $\zeta=\frac{3}{2}\cdot 1.5^{\circ}=2.25^{\circ}$. Within this angle the MS light grows to $\sim 9\%$, overestimating the number of particles at $X_{max}$ by this value. For shower parts near the ground, where shower age equals 1.25, this number grows to $\sim 27\%$. To find the shower primary energy $E_0$ (by integrating the energy deposit over depth $X_0$) it is necessary to extrapolate the shower for larger ages, so that the overestimation from these parts would be even larger. We estimate that not allowing for the MS effect would increase the reconstructed $E_0$ by  $\sim 15\%$.\\
Not allowing for the MS light would also lead to an overestimation of the depth of the shower maximum $X_{max}$. Fitting a new cascade curve to that from the previous example with the increased values according to the MS contribution, we obtain that the overestimation equals $\Delta X_{max}\simeq 35\,g\,cm^{-2}$. With the difference between the depths of an average proton and  iron  showers being $\sim 100\,g\,cm^{-2}$, this is not a negligible number for drawing conclusions about the mass composition of the highest energy cosmic rays. 
\subsection{Contribution of higher generations}
 At first sight it may seem odd that it is only the first two generations that contribute to the scattered light. One might expect that the main contribution should come from the generation with the number closest  to the value  $R/\lambda_{QD}$, according to the Poisson distribution (although we have shown that the first two generations scale  with $R/\lambda_Q $ rather than  with $R/\lambda_{QD}$).   For $\lambda_R = 18\, km$ and $\lambda_M = 15\, km$ and a shower close to the ground we have  $\lambda_{QD} = 8.2 \,km$ . Thus, if the distance $R = 24\, km$ then $R/\lambda_{QD}\simeq 3$, so should one expect the third and the fourth generations to contribute even more than the first one? The answer is no, since the scattered photons should fall into the detector field of view, so that it is not the path length of a scattering \emph{by any angle}  (what $\lambda_{QD}$ means ) that counts but by some \emph{particular values} of it.  This decreases the probability of such a scattering and can be viewed as an increase of  the (actually) effective mean free path length $\lambda_{eff}$ for the process \emph{we are interested in} (i.e. photons arriving at the detector). One could estimate  its value from the ratio $n_1/n_0$ or $n_2/n_0$ adopting for this aim a Poisson distribution of the number of  contributing  generations. The two obtained values differ typically by 30-60$\%$, so that an estimation of the share of the remaining generations on the basis of this uncertain $\lambda_{eff}$  is uncertain as well. However,  let us take  an example: $R=25\, km$, elevation angle $15^\circ$, $\zeta =2^\circ$, what are typical conditions for registered showers in the Auger Observatory. The two  values of the mean number of (effective) scatterings are $0.034$ (from $n_1/n_0$) and $0.061$ (from $n_2/n_0$). Assuming the larger one, one gets $2.7\%$ for a fraction of the third and higher generations.\\
However, as our earlier work  for a uniform medium \cite{giller}b)  shows, where the third generation was calculated, its share was two times smaller than that deduced from the Poisson model and the ratios $n_1/n_0$ and $n_2/n_0$, so that this estimation may be too high.
Thus, we estimate that the contribution of the higher number generations should be small, at least for the conditions of Auger. 

\subsection{Comparison with other work}
 \begin{figure}[th]
  \centering
  \includegraphics[height=3in, angle=270]{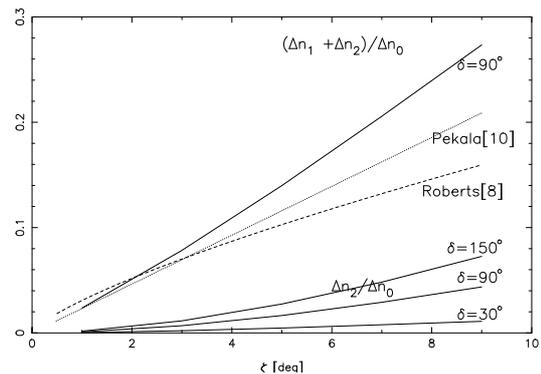}
  \caption{Ratio of the scattered to direct light $\Delta n/\Delta n_0$ integrated within angle $\zeta$, for shower 24 $km$ away from the detector, at elevation angle $10^{\circ}$.
Our results (first and second generation) are represented by thick solid lines for $\delta = 90^\circ$ and thin solid lines - for $\delta=30^{\circ}$ and $150^{\circ}$ (second generation). Parametrisation from Monte-Carlo works: Roberts - dashed line and P\c ekala et al - dotted line.}
  \label{fig18}
\end{figure}

The problem of the influence of the scattered light on the shower image in the direct light was first undertaken by Roberts \cite{roberts}. He treated it by Monte-Carlo simulations of tracks of individual photons emitted by vertical showers. Such shower (although, strictly speaking, non-existing) enabled the author to introduce a detector in a form of a ring around the shower as its symmetry axis, increasing dramatically the number of registered photons. Although the author appreciate that the singly scattered photons dominated the total MS signal, his simulations went to more than five scatters, including both Rayleigh and Mie. He obtained a parametrisation of the ratio of the MS signal to the total signal (MS + direct) in the form
\begin{equation}
K=77.4(OD\times \alpha\cdot R^{1/2}\cdot\zeta^{1.1})^{0.68}
\end{equation} 
where $OD$ is the total optical depth (molecular and aerosol) between the detector and the light source, $\alpha$ - total scattering coefficient (in $m^{-1}$) at the source. In our terms $OD=R/\lambda_{QD}$, 
$\alpha=\frac{1}{\lambda_Q}=\frac{1}{\lambda^R_Q}+\frac{1}{\lambda^M_Q}$, where $\lambda_{QD}$ is determined by formula (36) and $\lambda_Q$ depends only on the height in the stable atmosphere, so that (74) would mean the following
\begin{equation}
K\sim R\cdot\zeta^{0.75}\Big(\frac{1}{\lambda_{QD}\lambda_Q}\Big)^{0.68}
\end{equation} 
This parametrisation differs from those obtained in the present work although not very much. 
It has a weaker dependence on the angle $\zeta$, which in our case is as (about) $\sim \zeta^1$. The smaller power index at $\lambda_Q^{-0.68}$ ($\sim \lambda_Q^{-1}$ in our case) is to some extent compensated by the term $\lambda_{QD}^{-0.68}$.\\
Another work, dealing with the present problem was that of P\c{e}kala et al \cite{pekala}. The calculation method was Monte-Carlo as well. This paper did consider inclined showers so that the cylindrical symmetry, as in the previous case did not apply. Instead large "packages" of photons were assumed to behave in the same way, according to simulated points and directions of their scattering. After extensive simulations the authors arrived at a parametrisation of the ratio  $M$ of the MS signal to the direct one as follows
\begin{equation}
M\propto \frac{R}{\lambda_{QD}}\zeta \, e^{-\frac{h}{G}}
\end{equation}
with $G=5.43\,km$. For a one component exponential atmosphere with the scale height $G$ we would have
\begin{equation}
e^{-\frac{h}{G}}\propto\frac{1}{\lambda_Q}
\end{equation}
Indeed, it is not difficult to check that an effective scale height of the two-component atmosphere used by the authors for 
$2\le h\le 6 \,km$ is about $5\,km$ what approximately equals $G$. Thus, their parametrisation would be as 
\begin{equation}
M=\propto R\zeta\frac{1}{\lambda_{QD}\lambda_Q}=k_Q\zeta\frac{1}{\lambda_{QD}}
\end{equation}
It is not far from our derivation for the first generation, where the term $\frac{1}{\lambda_{QD}}$ does not appear. 
A comparison of the ratio of the total MS signal to the direct one as a function of $\zeta$, obtained in this study with that of Roberts and
P\c{e}kala et al is presented in Fig. 18. The agreement is quite good for $\zeta\le 4^\circ$ but for larger $\zeta$ the other authors seem to have underestimated a little the effect.  

\section{Summary and conclusions}
Previous studies of the influence of the scattering of photons on their way from the shower to the detector on 
the shower image used Monte-Carlo simulations. The effect has been called "the multiple scattering" since it was not clear how many times the photons arriving simultaneously with the direct light  actually scatter. Although Roberts \cite{roberts} appreciated the importance of photons scattered one time only he went on with simulating till several scatterings.\\
The main idea of this study was to divide the scattered light into separate generations: the first generation being photons scattered exactly one time, the second one two times, and so on. This enabled us to find analytically the angular distributions of photons as a function of time for the first generation which turned out to be the most important one. We also showed that any next generation can be found by some numerical integrations of the previous one. These results were obtained for a point source in a uniform medium as well as  in the actual atmosphere (although more numerical integrations were needed in the latter case). By considering separately the first generation it was easy to understand the differences between the two cases. The ratio of the second to the first generation is very small for small times after the light emission which will turn out to be relevant when considering the scattered light from extensive air showers what is a final aim of this study .\\ 
In particular, this aim was to calculate the angular distributions of scattered photons arriving simultaneously with the direct (not scattered) light emitted by an extensive air shower and find relevant parameters determining these distributions. The solutions for a point source were useful here since a shower could be treated as a series of consecutive (in space and time) point sources. For the first and the second generations (the most important) we have explained where actually the scattering points took place, so that we were able to integrate over all of them.\\
We have derived an analytical expression (although approximate) for the angular distribution  of the first generation photons, scattered by the Rayleigh process,  fitting quite well the exact numerical distributions. The analytical solutions have led us to a proper choice of the independent variables for a final parametrisation of the numerical results. For example the number of photons in the n-th generation turns out to be  proportional to  $(R/\lambda_Q )^n$, at least for $n\le 3$ (although for Mie with some correction).
Knowing the positions of the scatterings we were able to show that there should be a dependence on the viewing angle $\delta$, not taken into account in the considerations of the previous authors. Moreover, this dependence should be different for the Rayleigh and the Mie scatterings. Thus, we parametrised the first generation separately for these two processes. Concerning the second generation we combined all four possibilities of the two scatterings (RR, RM, MR and MM) to give just one parametrisation since  its  contribution was rather small. \\
From our parametrisation it follows that for $R\zeta = const$ (corresponding to a constant observed lateral dimension of a shower) the contribution of the MS component to the direct light depends practically only on the height of the observed shower element and on the viewing angle (the latter dependence being rather mild).
A typical overestimation of  the primary shower energy (if the MS effect was ignored) would be $\sim 15\%$ and of the depth of shower maximum $\Delta X_{max}\simeq 35\,g\,cm^{-2}$.
Our analytical and numerical approach gives similar results
as the previous Monte-Carlo work of other authors, although the parametrisation obtained in this paper is different. Only for larger opening angles $(\zeta > 4^{\circ})$ the prediction of the MS contribution obtained here is a bit larger, what would have some implications for a reconstruction of very close showers.\\
Summarising we conclude that the MS effect is not to be neglected while reconstructing parameters of the extensive air showers. To be properly allowed for a good knowledge of the atmospheric conditions is necessary
\cite{n} for determining values of $\lambda_Q$, as the shower traverses the atmosphere.\\

\paragraph{Acknowledgements}
This work has been financially supported by Polish Ministry of Science and Higher Education, grant No N N202 200239. It was stimulated by our participation in the Pierre Auger Collaboration. We thank the Collaboration for fruitful discussions.

\appendix
\section{Isotropic scattering of light in a uniform medium}
It may be of some value to know how much the fluxes $j^R(\theta,t;R)$ for the Rayleigh scattering differ from those when the scattering function $f(\alpha)=const$, so when the scattering is isotropic. To calculate it  we proceed as in Section 2 putting $f(\alpha)=\frac{1}{4\pi}$ in Eq. 6. We obtain
\begin{eqnarray}
j_1^{is}(\theta,t;R)=\frac{c}{8\pi^2\lambda\,R^2}\cdot\frac{e^{-k\tau}}{\tau^2-2\tau cos\theta+1}
\end{eqnarray}
and
\begin{eqnarray}
\lefteqn{\frac{d N_1^{is}(\zeta;t)}{dt}=\frac{c\,e^{-k\tau}}{16\pi\lambda R^2 \tau^2} \cdot} \\
&&\cdot\Big[(\tau^2+1)ln\Big(\frac{\tau^2-2\tau cos\zeta+1}{(\tau-1)^2}\Big) -2\tau(1-cos\zeta)\Big] \nonumber
\end{eqnarray}
The second generation can be obtained by proceeding as in Section 2.2, giving the result 
\begin{eqnarray}
j_2^{is}(\theta,t;R)=\frac{c\, e^{-k\tau}}{16\pi^2\lambda^2}\int_0^{x'_{max}}\frac{1}{\tau'}\,ln\frac{\tau'+1}{\tau'-1}\,
\frac{dx'}{R'^2}
\end{eqnarray} 
where $\tau'=\frac{ct-x'}{R'}$; $R'=\sqrt{x'^2-2Rcos\theta x'+R^2}$ and $x'_{max}=ct-R'=\frac{R}{2}\,\frac{\tau^2-1}{\tau-cos\theta}$.
To find 
$\frac{dN_2^{is}(\zeta; t)}{dt}$
one must integrate the flux over $\theta$ numerically.
How does the ratio of the two generations depend on the distance ? We can represent the first
 generation (A.2) in the  form
\begin{equation}
\frac{d N_1^{is}(\zeta, t)}{dt}\propto \frac{k}{R^3}f_1(t; \zeta)
\end{equation}
where $f_1$ depends on $t$ and $\zeta$ only.\\
The flux of the second generation  equals
\begin{eqnarray}
\lefteqn{j^{is}_2\propto \frac{k^2}{R^3}e^{-k\tau}\cdot} \\
&&\cdot\int_0^{u'_{max}(\tau, \theta)}   \frac{1}{\tau'}\,ln\frac{\tau'+1}{\tau'-1}\cdot\frac{du'}{u'^2-2u'cos\theta+1} \nonumber
\end{eqnarray}
where $u'=\frac{x'}{R}$, $u'_{max}=\frac{1}{2}\,\frac{\tau^2-1}{\tau-cos\theta}$, 
$\tau'=\frac{\tau-u'}{\sqrt{u'^2-2u'cos\theta+1}}$. \\
We see that the integral depends on $\tau$ and $\theta$ only so that
\begin{equation}
\frac{dN_2^{is}}{dt}\propto \frac{k^2}{R^3}f_2(\tau; \zeta)
\end{equation}
Thus
\begin{equation}
\frac{dN^{is}_2}{dt}/\frac{d N^{is}_1}{dt} \propto k \,f(\tau; \zeta)
\end{equation}
So that for a given time the ratio is proportional to the distance $R$ 
measured in units of $\lambda$. It can be verified that this relation holds for any 
scattering process in a uniform medium, not only for isotropy.

\section{Derivation of an analytical expression of the angular distribution of 
the first  generation $d M_1^R/d\theta$ for small angles $\theta$}

The formula for the angular distribution of the  first generation integrated over time, can be rewritten in the following form
\begin{eqnarray}
\lefteqn{\frac{d M_1}{d\theta}=\frac{k}{2}\frac{|cos\theta|}{R^2} \cdot} \\
&&\cdot\int_0^{\pi-\theta}exp\Big[-k\frac{sin\beta+sin\theta}{sin(\beta+\theta)}\Big]f(\beta+\theta)d\beta \nonumber
\end{eqnarray}
where $f(\beta+\theta)=f(\alpha)$ is the angular distribution of the scattering angle $\alpha$,
normalised to $1$ when integrated over the full solid angle.\\
For the Rayleigh scattering we obtain
\begin{eqnarray}
\lefteqn{\frac{d M_1^R}{d\theta}=\frac{3k}{32\pi}\frac{|cos\theta|}{R^2}\cdot} \\
&&\cdot\int_0^{\pi-\theta} exp\Big[-k\frac{sin\beta+sin\theta}{sin(\beta+\theta)}\Big]\Big[1+cos^2(\beta+\theta)\Big]
d\beta \nonumber
\end{eqnarray}
To find the above integral for $\theta\ll 1$ we divide the integration region into two parts:
\begin{eqnarray}
\lefteqn{\frac{d M_1^R}{d\theta}=C\Big[\int_0^{\pi-\theta-\epsilon}
...d\beta+\int_{\pi-\theta-\epsilon}^{\pi-\theta}...d\beta\Big]= } \nonumber \\
&&= C(I_1+I_2)
\end{eqnarray}
where $C=\frac{3k}{32\pi}\frac{|cos\theta|}{R^2}$.\\
Choosing $\theta\ll \epsilon\ll 1$  it will be possible to find  analytically the integrals $I_1$ 
and $I_2$. Each of them is a sum of the two integrals:
\begin{equation}
\begin{array}{l}
\displaystyle I_i=\int_{\beta_{i1}}^{\beta_{i2}}exp\Big[-k\frac{sin\beta+sin\theta}{sin(\beta+\theta)}\Big]d\beta+ \\
\displaystyle +\int_{\beta_{i1}}^{\beta_{i2}} exp\Big[-k\frac{sin\beta+sin\theta}{sin(\beta+\theta)}\Big]cos^2(\beta+\theta)d\beta=\\
\displaystyle =I_{i1}+I_{i2}
\end{array}
\end{equation}
for $i=1, \,2$ and $\beta_{11}=0$, $\beta_{12}=\beta_{21}=\pi-\theta-\epsilon$, $\beta_{22}=\pi-\theta$.
\\
Using $\theta\ll 1$ we have
\begin{eqnarray}
I_{11}&\simeq &\int_0^{\pi-\theta-\epsilon} 
exp\Big[-k\frac{sin\beta(1+\frac{\theta}{sin\beta})}{sin\beta(1+\theta\, ctg\beta)}\Big]d\beta\simeq \nonumber \\
&\simeq & \int_0^{\pi-\theta-\epsilon} exp\big[-k(1+\frac{\theta}{sin\beta}-\theta\, ctg\beta)\big]d\beta =\nonumber \\
& = & e^{-k}\int_0^{\pi-\theta-\epsilon} exp\big(-k\,\theta \,tg\frac{\beta}{2}\big)d\beta
\end{eqnarray}
where we have neglected  all terms $\propto \theta^2$.
Note that in the integration region $\theta \,ctg \beta<\frac{\theta}{\epsilon}\ll 1$ so that
our approximation 
\begin{equation}
(1+\theta\, ctg \beta)^{-1}\simeq 1-\theta \,ctg\beta
\end{equation}
is justified. For the same reason the exponential index $k\theta\, tg\frac{\beta}{2}<k\,\theta\,\frac{2}{\epsilon}\ll 1$
so that
\begin{eqnarray}
I_{11}&\simeq&  e^{-k}\int_0^{\pi-\theta-\epsilon}  \Big(1-k\theta\, tg\frac{\beta}{2}\Big)   d\beta = \\
&&= e^{-k}\Big[\pi-\theta-\epsilon+2k\theta \,ln\,\Big(sin\frac{\epsilon+\theta}{2}\Big)\Big]\simeq \nonumber \\
&& \simeq  e^{-k}(\pi-\theta-\epsilon+2k\theta\,ln\frac{\epsilon}{2}) \nonumber
\end{eqnarray}
The integral $I_{12}$ can be found easily:
\begin{eqnarray}
I_{12}&\simeq & e^{-k}\int_0^{\pi-\theta-\epsilon}  \Big(1-k\theta\, \frac{1-cos\beta}{sin\beta}\Big)\cdot \\
&\cdot&(cos^2\beta-2\theta sin\beta cos\beta)   d\beta \simeq \nonumber \\
&\simeq &  e^{-k}\Big[\frac{\pi}{2}-\theta-\epsilon+2k\theta\,(ln\frac{\epsilon}{2}+1)\Big]\nonumber
\end{eqnarray}
Thus
\begin{eqnarray}
I_1&=&I_{11}+I_{12}= \\
&=& e^{-k}\Big[\frac{3}{2}\pi-2\theta-2\epsilon+2k\theta(2ln\frac{\epsilon}{2}+1)\Big]\nonumber
\end{eqnarray}
When calculating $I_2$ we introduce an angle $\chi=\pi-\theta-\beta$ to take advantage of $\chi$ being small in the integration region of $I_2$.\\
We have
\begin{eqnarray}
I_{21}& = & \int_0^{\epsilon}exp\Big\{-k\frac{\theta+(\theta+\chi)[1-\frac{(\theta+\chi)^2}{6}] }{\chi(1-\frac{\chi^2}{6})}   \Big\}d\chi\simeq \nonumber \\
&\simeq & e^{-k}\int_0^{\epsilon}exp\Bigg(-\frac{2k\theta}{x}\Bigg)dx=e^{-k}2k\theta\int_{\frac{2k\theta}{\epsilon}}^{\infty}\frac{e^{-y}dy}{y^2} \nonumber \\
&\simeq & e^{-k}\Big[2k\theta\big(C_{Eu}-ln\frac{\epsilon}{2}+ln\,k\theta -1\big)+\epsilon\Big]
\end{eqnarray}
where $C_{Eu}\simeq 0.577$ is the Euler constant.\\
The last integral, $I_{22}$, equals
\begin{eqnarray}
I_{22}& = & \int_0^{\epsilon}exp\Big\{-k\frac{2\theta+\chi}{\chi}(1-\chi^2)d\chi \simeq \\
&\simeq & e^{-k}\Big[2k\theta\int_{\frac{2k\theta}{\epsilon}}^{\infty}\frac{e^{-y}dy}{y^2}-(2k\theta)^3
\int_{\frac{2k\theta}{\epsilon}}^{\infty}\frac{e^{-y}dy}{y^4}\Big]\nonumber
\end{eqnarray}
The second integral is of the order of $\epsilon^2$, so that $I_{22}=I_{21}$ and
\begin{eqnarray}
I_2=e^{-k}\Big\{4k\theta[C_{Eu}-ln\frac{\epsilon}{2}+ln(k\theta)-1]+2\epsilon\Big\}\nonumber \\
\,
\end{eqnarray}
Thus
\begin{eqnarray}
I_{1}+I_{2}=e^{-k}\Big\{\frac{3}{2}\pi-2\theta+4k\theta[ln(k\theta)+C_{Eu}-\frac{1}{2}\big]\Big\}\nonumber \\
\,
\end{eqnarray}
and finally
\begin{eqnarray}
\frac{dM_1^R}{d\theta}=\frac{9ke^{-k}}{64R^2}
\Big[1-\frac{4\theta}{3\pi}+\frac{8k\theta}{3\pi}\big(ln(k\theta)+C_{Eu}-\frac{1}{2}\big)\Big]\nonumber \\
\,
\end{eqnarray}
As should be expected, the terms depending on $\epsilon$ have cancelled in the final result.\\
The ratio of the scattered light arriving within a certain angle $\zeta$ to the point source direction to that non-scattered (direct) equals
\begin{eqnarray}
\frac{M_1^R}{N_0}=\frac{9\pi}{16}k\Big\{\zeta+\frac{4k\zeta^2}{3\pi}\Big[
ln(k \zeta)+C_{Eu}-1-\frac{1}{2k}\Big]\Big\}\nonumber\\
\,
\end{eqnarray}

\section{Towards an analytical solution of the second generation in a uniform medium}
Here we present results of some analytical integrations of functions $F(\theta', \alpha)$ and $G(R',\alpha)$ defined in Section 2.2.
For the Rayleigh scattering we have
\begin{equation} 
f^R(\alpha')=\frac{3}{16\pi}(1+cos^2\alpha')
\end{equation} 
Since $cos\alpha'=cos\alpha\,cos\theta'-sin\alpha\,sin\theta'\,cos\phi'$, $F^R(\theta',\alpha)$ (Eq. 25) can be easily found analytically
\begin{eqnarray} 
&&F^R(\theta',\alpha)=\int^{2\pi}_0 f^R(\alpha')d\phi'= \\
&&=\frac{3}{8}\Big[1+\frac{1}{2}sin^2\alpha+(cos^2\alpha-\frac{1}{2}sin^2\alpha)cos^2\theta'\Big] \nonumber
\end{eqnarray} 
The calculation of a corresponding function $F^M(\theta', \alpha)$ for the Mie scattering is a little more complicated. The adopted here form of $f^M(\alpha')$ has been chosen in such a way as to make the integration over $\phi'$ analytically possible. We have (Eq. 16)
\begin{eqnarray} 
&&F^M(\theta',\alpha')=\int_0^{2\pi}f^M(\alpha')d\phi'= \\
&&=2\int_0^{\pi}(a_i\,cos^8\alpha'+b)d\phi'\nonumber
\end{eqnarray} 
The angle of the second scattering $\alpha'$ changes within limits $\alpha'_{min}\le\alpha'\le\alpha_{max}'$
where
\begin{eqnarray} 
\alpha'_{min}&=&|\theta'-\alpha| \nonumber \\
 \alpha'_{max}&=& \theta'+\alpha \quad \textrm{if} \quad \theta'+\alpha<\pi \nonumber \\
 \alpha'_{max}&=& 2\pi-(\theta'+\alpha) \quad \textrm{if} \quad \theta'+\alpha>\pi
\end{eqnarray} 
Since $\alpha'=\frac{\pi}{2}$ separates the two regions of $\alpha'$ where $f^M(\alpha')$ has different shapes, the solution of (C.3) depends on the positions of $\alpha'_{min}$ on the $\alpha'$ axis with respect to the point $\alpha'=\frac{\pi}{2}$. There are three cases to be considered \\
\paragraph{a). $\alpha'_{max} < \frac{\pi}{2}$}
Then, of course, $\alpha'_{min}<\frac{\pi}{2}$ as well and the function $f^M(\alpha')=f_1^M(\alpha')$ applies to the whole region of $\phi'$ ($0\le\phi'\le 2\pi $). Thus we have
  \begin{eqnarray} 
F^M &=& 2\Big[\int_0^{\pi}a_1\big(A+B\,cos\phi'\big)^8d\phi'+\pi b\Big] \nonumber \\
&=&2\Big[a_1\sum_{k=0}^8C_k\int_0^{\pi}
cos^k\phi'\,d\phi'+\pi b\Big]
\end{eqnarray}
where $C_k={8 \choose k}\,A^{8-k}B^k$, $A=cos\alpha\,cos\theta'$ and $B=sin\alpha\, sin\theta'$.\\
For odd $k$ the integrals in (C.5) vanish. For even $k=2m$ we have
\begin{equation} 
J_{2m}=\int_0^{\pi}cos^{2m}\phi'd\phi'=\frac{(2m-1)!!}{2m!!}\pi
\end{equation} 
Hence
\begin{equation} 
F^M=2\Big[a_1\sum_{m=0}^4 C_{2m}J_{2m}+\pi b\Big]
\end{equation}  
\paragraph{b). $\alpha'_{min}<\frac{\pi}{2}<\alpha'_{max}$} 

This case is a bit more complicated since there are both parts $f_1^M(\alpha')$ and $f_2^M(\alpha')$ involved. The value $\phi'_*$ corresponding to $\alpha'=\frac{\pi}{2}$ equals $cos\phi'_*=ctg\alpha\cdot ctg\theta'$.\\
We have 
  \begin{eqnarray} 
F^M &=&2\Big[ \int_0^{\phi'_*}f_1^M(\alpha')d\phi'+ \int_{\phi'_*}^{\pi}f_2^M(\alpha')d\phi'\Big]=  \nonumber\\
&= &2\sum_{k=0}^8C_k\Big[a_1\int_0^{\phi'_*}cos^k\phi'd\phi'+\nonumber \\
&+&a_2\int_{\phi'_*}^{\pi}cos^k\phi'd\phi'\Big]+2\pi b
\end{eqnarray}
Solving the integrals we finally obtain
  \begin{eqnarray} 
F^M &=& 2\Bigg[ \sum_{n=0}^4 C_{2n}\Big\{(a_1-a_2)\Big[ \frac{1}{2^{2n}}{2n \choose n}\phi'_*+G_n(\phi'_*)\Big]+\nonumber \\
&&+\frac{a_2}{2^{2n}}{2n \choose n}\pi\Big\} \sum_{n=0}^3 C_{2n+1}(a_1-a_2)G'_n(\phi'_*)+b\pi\Bigg]\nonumber\\
&&\quad
\end{eqnarray}
where
  \begin{eqnarray} 
G_n(\phi'_*) &=& \frac{1}{2^{2n-1}}\sum_{i=0}^{n-1}{2n \choose i}\frac{sin[(2n-2i)\phi'_*]}{2n-2i}  \\
G'_n(\phi'_*) &=& \frac{1}{2^{2n}}\sum_{i=0}^{n}{2n+1 \choose i}\frac{sin[(2n-2i+1)\phi'_*]}{2n-2i+1}\nonumber
\end{eqnarray}
\paragraph{c). $\alpha'_{min}>\frac{\pi}{2}$ }
Then, of course  $\alpha'_{min}>\frac{\pi}{2}$ as well and for all values of $\phi'$ we have that $\alpha'>\frac{\pi}{2}$, so that $f_1^M(\alpha')=f_2^M(\alpha')$. The calculation of $F^M$ is the same as in case a) but $a_1\to a_2$. Thus
\begin{equation} 
F^M=2\Big[a_2\sum_{m=0}^4 C_{2m}J_{2m}+\pi b\Big]
\end{equation} 
The next integration over $\theta'$ (Eq. 26) defining $G^{AB}(R',\alpha)$ can be found analytically only\footnote{actually for $G^{MR}$, $G^{RM}$ and $G^{MM}$ the result could also be found analytically but due to the high power of $cos\alpha$ (index = 8) there would be very many terms making such an effort worthless.} for $G^{RR}$. The calculation is lengthily so that we will present here only the final result for the flux of the second generation 
\begin{eqnarray} 
j_2^{RR}(\theta, t)&&=\frac{1}{2}\Big(\frac{3}{8\pi}  \Big)^2 \frac{c}{\lambda_R^2}e^{-\frac{ct}{\lambda}}\int_0^{x'_{max}}I(\tau',\alpha)\frac{dx'}{R'^2}\nonumber \\
&&\quad
\end{eqnarray} 
where 

\begin{eqnarray} 
&&I(\tau', \alpha) \equiv  \int_0^{\pi} \frac{1}{y'}\Big(1-\frac{2sin^2\theta'}{y'}+\frac{2sin^4\theta'}{y'^2}\Big)\cdot\nonumber\\
&&\cdot\Big(a+b\,cos^2\theta'\Big)sin\theta'\,d\theta'= \\
&=& \frac{1}{16\tau'^2}\times \Bigg\{\Big[a+\frac{b(\tau'^2+1)^2}{4\tau'^2}\Big]\cdot\Big(\sum^2_{i=-2 \atop i\neq 0}
a_i\frac{x_2^i-x_1^i}{i}+\nonumber \\
&+&a_0\, ln\frac{x_2}{x_1}+ \frac{b}{4\tau'^2}\Bigg[\sum_{i=1}^4 a_{i-2}\cdot\frac{x_2^i-x_1^i}{i}+a_{-2}ln\frac{x_2}{x_1}+\nonumber\\
&-&2(\tau'^2+1)\cdot  \Big(\sum^3_{i=-1 \atop i \neq 0}a_{i-1}\frac{x_2^i-x_1^i}{i}+a_{-1}ln\frac{x_2}{x_1}  \Big) \Bigg]\Bigg\} \nonumber
\end{eqnarray}
The coefficients $a_i$ are the same as in (13); $x_1=(\tau'-1)^2$, $x_2=(\tau'+1)^2$, $a=1+\frac{1}{2}sin^2\alpha$, $b=cos^2\alpha-\frac{1}{2}sin^2\alpha$. The integral over $x'$ has to be found numerically.


\begin{thebibliography}{00}
\bibitem{hires} T. Abu-Zayyad et. al,  Astroparticle Physics 12 (1999) 121-134
\bibitem{pao} The Pierre Auger Collaboration: J. Abraham et al, Nucl. Instrum. Meth. A 620 (2010) 227-251
\bibitem{tarr} T. Nonaka et al, Nuclear Physics B Proc. Suppl. 190 (2009) 26-31 
\bibitem{dedx} The MACFLY Collaboration: P. Colin et al,  Astroparticle Physics 30 (2009) 312-317
\bibitem{espec} M. Giller et al, J. Phys. G: Nucl. Part. Phys. 30 (2004) 97-105
\bibitem{simil} M. Giller et al, J. Phys. G: Nucl. Part. Phys. 31 (2005) 947-958
\bibitem{nerl} F. Nerling at al, Astroparticle Physics 24 (2006) 421
\bibitem{latch} M. Giller  and G. Wieczorek, Astroparticle Physics 31 (2009) 212-219
\bibitem{roberts} M. D. Roberts, J. Phys. G: Nucl. Part. Phys. 31 (2005) 1291-1301
\bibitem{giller}
a) M. Giller and A. \'Smia\l{}kowski, (2005) Proceedings of the 29th Int. Cosmic Ray Conference, Pune, India    \\    
b) M. Giller and A. \'Smia\l{}kowski, (2007) Proceedings of the 30th Int. Cosmic Ray Conference, Merida, Mexico\\ 
c) M. Giller and A. \'Smia\l{}kowski, (2011) Proceedings of the 32nd Int. Cosmic Ray Conference, Beijing, China

\bibitem{pekala} J. P\c ekala et al,  Nucl. Instrum. Meth. A605 (2009) 388-398
\bibitem{somm} P. Sommers,  Astroparticle Physics 3 (1995) 349-360
\bibitem{n} The Pierre Auger Collaboration: J. Abraham et al,  Astroparticle Physics 33 (2010) 108-129
\end{thebibliography}

\end{document}